\documentclass[%
 reprint,
superscriptaddress,
 amsmath,amssymb,
 aps,
pra,
floatfix,
longbibliography
]{revtex4-1}

\usepackage{ifpdf}
\usepackage{amsmath} 
\usepackage{amssymb}
\usepackage{amsfonts}
\usepackage{braket}
\usepackage{color}
\usepackage[colorlinks=true,linkcolor=blue,citecolor=blue,breaklinks]{hyperref}
\usepackage{graphicx}
\usepackage{dcolumn}
\usepackage{bm}
\usepackage[normalem]{ulem}
\usepackage{verbatim}

\newcommand{\hc}{\text{H.c.}}
\newcommand{\la}{\langle}
\newcommand{\ra}{\rangle}

\newcommand{\hH}{\hat{H}}
\newcommand{\hV}{\hat{V}}

\definecolor{Zcolour}{rgb}{0.992, 0.588, 0.22}
\definecolor{purple}{rgb}{0.5,0,0.5}
\definecolor{brown}{rgb}{0.6,0.2,0}
\definecolor{dkgreen}{rgb}{0,0.5,0}

\begin{document}


\title{Slow thermalization of exact quantum many-body scar states under perturbations}
\author{Cheng-Ju Lin}
\affiliation{Perimeter Institute for Theoretical Physics, Waterloo, Ontario, Canada, N2L 2Y5}
\author{Anushya Chandran}
\affiliation{Department of Physics, Boston University, Boston, MA 02215, USA}
\author{Olexei I.~Motrunich}
\affiliation{Department of Physics, California Institute of Technology, Pasadena, CA 91125, USA}

\date{\today}

\begin{abstract}
Quantum many-body scar states are exceptional finite energy density eigenstates in an otherwise
thermalizing system that do not satisfy the eigenstate thermalization hypothesis.
We investigate the fate of exact many-body scar states under perturbations.
At small system sizes, deformed scar states described by perturbation theory survive.
However, we argue for their eventual thermalization in the thermodynamic limit from the finite-size scaling of the off-diagonal matrix elements.
Nevertheless, we show numerically and analytically that the nonthermal properties of the scars survive for a parametrically long time in quench experiments. 
We present a rigorous argument that lower-bounds the thermalization time for any scar state as $t^{*} \sim O(\lambda^{-1/(1+d)})$, where $d$ is the spatial dimension of the system and $\lambda$ is the perturbation strength.
\end{abstract}

\maketitle

\section{Introduction}
Wigner pioneered the application of random matrix theory to describe quantum chaos \cite{mehtamadanlalRandom}.
Nowadays, it has even become a definition of quantum chaos: the applicability of the random matrix theory description to the statistical properties of the spectrum and wavefunctions of a quantum mechanical system, in both single-particle and many-body quantum systems.
Based on the random matrix theory description, Srednicki and Deutch proposed the \textit{Eigenstate Thermalization Hypothesis} (ETH), bridging the concept of quantum chaos and the validity of statistical mechanics in closed quantum many-body systems \cite{deutschQuantum1991,srednickiChaos1994}.
Essentially, ETH implies that the reduced density matrix of a single eigenstate is equal to that of the microcanonical/canonical ensemble, and it is how statistical mechanics emerges in closed quantum systems.

The strong version of ETH proposes that \textit{every} eigenstate satisfies the above property~\cite{rigolRelaxation2007,rigolThermalization2008,kimTesting2014,PhysRevX.8.021026}.
However, there can be some exceptional states at a finite energy density that do not satisfy the ETH, while the other states do.
Such states are dubbed \textit{quantum many-body scar states}, in analogy with the single-particle scar states~\cite{hellerBoundState1984}.
Recently, there has been a surge in interest of finding and understanding quantum many-body scar states due to the observation of anomalous dynamics in a Rydberg atom experiment~\cite{bernienProbing2017}.
Known systems that host quantum many-body scar states include the PXP model describing the Rydberg-blockaded atom chain~\cite{turnerWeak2018,turnerQuantum2018,khemaniSignatures2019,schecterManybody2018,linExact2019,iadecolaQuantum,suraceLattice,Shiraishi_2019}, the Affleck-Kennedy-Lieb-Tasaki model \cite{affleckRigorous1987,moudgalyaExact2018,moudgalyaEntanglement2018} and the spin-1 XY model~\cite{schecterWeak2019}.
References~\cite{shiraishiSystematic2017,moriThermalization2017} developed a systematic construction to embed nonthermal states in the spectrum.
Many other systems or models have also been discovered or constructed to have scars or scar-like physics~\cite{kormosRealtime2016,vafekEntanglement2017,robinsonSignatures2019,jamesNonthermal2019,choiEmergent2019,okTopological,michailidisSlow,bullSystematic2019,moudgalyaQuantum,khemaniLocal,paiDynamical2019,cuberoLack}.

Current analytical understanding about quantum many-body scars relies on the identification of certain exact eigenstates. 
Moreover, the Hamiltonians with exact scar states are at some special point in some parameter space or have the embedded Hamiltonian structure.
An immediate question arises: how robust are the exact quantum many-body scar states under generic perturbations?
Are the exact scar states discovered and constructed in several models useful to understand the physics once we add perturbations? 

In this paper, we address the above questions by studying the fate of the exact scar states under perturbations.
While we use the perturbed PXP model as our main showcasing example, our arguments are in fact general and apply to any model with exact scars that is subjected to generic perturbations.
For some analysis, we also study the perturbed spin-1 XY model to further support our arguments.

We first show that, in the finite-size ED data, there is some apparent robustness in the nonthermal signatures of the exact scar states upon perturbation.
The perturbed eigenstates in finite sizes can indeed be understood using standard perturbation theory.
However, the finite-size scaling of the matrix elements between the scar states and other eigenstates suggests the eventual thermalization of the scar states at larger system sizes.
In both studied models, the scaling of the matrix elements between the scar and thermal states is well described using the random-matrix theory picture of the thermal states. This predicts that such matrix elements scale as ${\mathcal D}^{-1/2}$, where ${\mathcal D}$ is the many-body Hilbert space dimension (of the relevant symmetry sector) and grows exponentially with the system volume.  The ${\mathcal D}^{-1/2}$ dependence follows solely from the property of the thermal states, while only the numerical amplitude depends on details of the scar states.  Since the many-body level spacing decreases much faster as ${\mathcal D}^{-1}$, this very general mixing argument suggests an eventual demise of exact scar eigenstates in any system that is subjected to generic perturbations.

Despite the eventual thermalization of the scar states, the thermalization rate of observables is parametrically slow in the strength of the perturbation.
In particular, we consider the quench dynamics starting from the exact scar states evolving under the perturbed Hamiltonian. 
We rigorously lower-bound the time scale for the expectation value of any local observable to thermalize by $t^* \sim O(\lambda^{-\frac{1}{1+d}})$, where $\lambda$ is the strength of the perturbation and $d$ is the spatial dimension of the system.
The long protection of the nonthermal property is due to the locality of the Hamiltonian.
Roughly, in the Heisenberg picture, the time derivative of the expectation value of a local observable is given by the expectation value of the commutator of the perturbed Hamiltonian and the time-evolved observable.
Writing the perturbation as a sum of local terms, the contributions from the locations outside of the ``light cone" relative to the observable location are controlled by the Lieb-Robinson bound, while each contribution from inside the light cone is bounded by an $O(\lambda)$ number.
This gives that the expectation value in time can only deviate from its initial nonthermal value by no greater than $\lambda (c_0 t^d + c_1 t^{1+d})$, leading to the above lower bound on the thermalization time.
The bound also applies to the survival time of any anomalous dynamics such as persistent oscillations in some models with a tower of equally spaced scar states~\cite{moudgalyaExact2018, choiEmergent2019, schecterWeak2019}.
While this bound is likely not optimal (e.g., our numerical study in the perturbed PXP model suggests that $t^*$ diverges at least as strongly as $1/\lambda$; while the numerical study in the perturbed spin-1 XY model suggests $t^* \sim 1/\lambda ^{2}$), it is completely rigorous and general.
A consequence of our results is that the exact scar states discovered or constructed in different special models can be used to understand the persisting dynamical signatures under perturbations up to some parametrically large time scale and in the thermodynamic limit. 

Our results suggest some analogy between weakly perturbed scarred systems and weakly perturbed integrable systems. 
In the latter case, the eigenstates in finite sizes can be understood as perturbed from the special integrable Hamiltonian.
Analogously, the deformed scar states in finite sizes are perturbatively related to the exact scar states of the corresponding special Hamiltonian. 
Moreover, in the thermodynamic limit, a weakly perturbed integrable Hamiltonian prethermalizes to a generalized Gibbs ensemble that persists for a parametrically long time scale~\cite{bertiniPrethermalization2015, bertiniThermalization2016, mallayyaPrethermalization2019}, believed to diverge as $O(\lambda^{-2})$.
Similarly, the nonthermal properties of the exact scar states can also survive in quench experiments under perturbations to some parametrically long time even in the thermodynamic limit, which we also expect to diverge as power law in $\lambda$ in generic cases.
We therefore propose that there is a similarity at this level between ``completely solvable" (integrable) Hamiltonians and ``partially solvable" scar Hamiltonians.
We note that by the latter term we mean that only some special eigenstates are known analytically, while the other eigenstates are not known analytically and are in overwhelming numbers thermal.
Correspondingly, it is only these special eigenstates or the corresponding special initial states that produce observable scar signatures either in small system sizes or quench dynamics. This aspect is different from the integrable systems where all eigenstates are special and essentially any initial state will show prethermalization.
We finally remark that our rigorous bound on the thermalization time does not use Fermi golden rule type arguments that give $O(\lambda^{-2})$ prethermalization time for nearly integrable systems or when a conservation law is weakly broken~\cite{bertiniPrethermalization2015, bertiniThermalization2016, mallayyaPrethermalization2019}. 
It is an open question whether such arguments can be extended to the scar thermalization problem.

The paper is organized as follows.
In Sec.~\ref{sec:model}, we introduce the PXP model and its properties, and also motivate the specific type of perturbation in our consideration.
In Sec.~\ref{sec:signatures}, we study the signatures of the scars under the perturbation in finite size systems, demonstrating their apparent robustness.
In Sec.~\ref{subsec:perturb}, we apply standard perturbation theory to describe the deformed scar states at small perturbation strength and numerically accessible system sizes. 
However, in Sec.~\ref{subsec:finitesizescaling}, we argue that the deformed scar states eventually hybridize with the ETH-satisfying states at nearby energies by studying the finite-size scaling of the matrix elements connecting the scar states to other states. Thus, the deformed scar states lose their nonthermal character as $L \to \infty$ and eventually satisfy the ETH.
Despite their eventual thermalization, in Sec.~\ref{subsec:tebd}, we numerically show that the thermalization is slow under global quenches, and prove a rigorous lower bound on the thermalization time scale in Sec.~\ref{subsec:rigorousbound}.
In Sec.~\ref{subsec:strongerbound}, we discuss possible scenarios where the bound on the thermalization time can be stronger.
We conclude and discuss our work in Sec.~\ref{sec:conclusions}.
Appendices~\ref{app:summaryexactscar}-\ref{app:distr_matr_elem} contain more details of the perturbed PXP model study (including general discussion of some properties of the distribution of matrix elements dictated by the locality of the Hamiltonian), while Appendix~\ref{app:spin-1XY} presents both the matrix element and quench dynamics study of the perturbed spin-1 XY scar model.

\section{Model}\label{sec:model}
As a specific example for the numerical study, we consider the one-dimensional (1D) PXP model with perturbation.
(An additional numerical study on the perturbed spin-1 XY model is organized in Appendix~\ref{app:spin-1XY}.)
The PXP model is the effective constrained model describing the dynamics of the Rydberg atom chain in the regime of the nearest-neighbor blockade.
More specifically, we consider a 1D atom chain with size $L$ and open boundary conditions.
There are two degrees of freedom at each site: $|0\ra$ (atomic ground state), and $|1\ra$ (atomic excitation).
The blockade condition excludes all configurations $|\dots 11 \dots \ra$ with adjacent atomic excitations from the Hilbert space. 
Despite the non-tensor product structure, one can still have the concept of ETH in the constrained Hilbert space \cite{chandranEigenstate2016}.
The dimension of the Hilbert space grows as $\mathcal{D}_L \sim \phi^L$, where $\phi=(1+\sqrt{5})/2$ is the golden ratio.

The perturbed PXP model has the following Hamiltonian: 
\begin{equation}
    \hH = \hH_0 + \lambda \hV ~,
\end{equation}
where
\begin{equation}\label{eqn:PXP}
    \hH_0 = X_1 P_2 + \sum_{j=2}^{L-1} P_{j-1} X_j P_{j+1} + P_{L-1} X_L ~. 
\end{equation}
Here $P \equiv |0\ra \la 0|$ and $X \equiv |0\ra \la 1| + |1\ra \la 0|$.
Before specifying $\hV$, we summarize some properties of $\hH_0$, in order to motivate the specific perturbation we will consider later.
These properties of the PXP model have been discussed in detail in Refs.~\cite{turnerWeak2018,turnerQuantum2018,schecterManybody2018,khemaniSignatures2019,linExact2019}.

First, $\hH_0$ has the inversion symmetry $\hat{I}: j \rightarrow L-j+1$.
It also has the property that if we define the particle-hole transformation 
\begin{equation}
   \hat{\mathcal{C}} = \prod_j Z_j ~,
\end{equation}
where $Z \equiv |1\ra \la 1| - |0\ra \la 0|$, then $\hat{\mathcal{C}} \hH_0 \hat{\mathcal{C}} = - \hH_0$.
This implies that, for any eigenstate $|E\ra$ of $\hH_0$ with energy $E \neq 0$, $\hat{\mathcal{C}}|E\ra$ is also an eigenstate with energy $(-E)$.

The combination of $\hat{I}$ and $\hat{\mathcal{C}}$ guarantees the exponential degeneracy of $E=0$ states in $\hH_0$.
This is due to the difference between the number of states with $\mathcal{C}=1$ and $\mathcal{C}=-1$ in each inversion symmetry sector $I$.
As a result, in the $E=0$ manifold, the eigenstates with a particular inversion symmetry quantum number $I = \pm 1$ will also have a definite particle-hole quantum number given by $\mathcal{C} = I$.

\subsection{Signatures of the exact scar states}
In Ref.~\cite{linExact2019}, four exact scar states were discovered in $\hH_0$ for even system sizes $L$, labeled as $|\Gamma_{\alpha\beta} \ra$, where $\alpha, \beta \in \{1, 2\}$.
Here we assume that the states are normalized.
For the readers' convenience, we summarize the wavefunctions and some essential properties of these states in Appendix~\ref{app:summaryexactscar}.
In particular, $|\Gamma_{11} \ra$ and $|\Gamma_{22} \ra$ have energies $E=0$; the states $|\Gamma_{12} \ra$ and $|\Gamma_{21} \ra$ have energies $E=\sqrt{2}$ and $E=-\sqrt{2}$ respectively.

Despite being in the middle of the spectrum at energy density corresponding to infinite temperature, the states $|\Gamma_{\alpha\beta} \ra$ have constant bipartite entanglement entropy scaling with the subsystem length (``area law"), instead of the volume law scaling required by the ETH. 
Moreover, the expectation values of some local observables in these states do not agree with the thermal ensemble values at infinite temperature, therefore violating the ETH.
One intriguing feature is their valence bond solid (VBS) order, which is also used to identify their translation symmetry breaking.
The order parameter is defined as 
\begin{equation}
    \hat{M} \equiv \frac{1}{L-1} \sum_{j=1}^{L-1} (-1)^j \hat{D}_{j,j+1} ~,
    \label{eqn:MVBS}
\end{equation}
where
\begin{equation}
    \hat{D}_{j,j+1} \equiv |01 \ra \la 10| + \hc
\end{equation}
detects the dimer (bond) strength.
In the thermodynamic limit, in the bulk,  $\la\Gamma_{\alpha\beta}|\hat{D}_{j,j+1}|\Gamma_{\alpha\beta}\ra = 0 $ if $j$ is odd and $-2/9$ if $j$ is even.
On the other hand, an infinite temperature thermal ensemble would give the thermal value $\frac{1}{\mathcal{D}_L}\text{Tr} [ \hat{D}_{j,j+1}] = 0$ for all $j$, where $\mathcal{D}_L$ is the dimension of the Hilbert space of system size $L$.

\subsection{Particle-hole odd perturbation}
We are interested in the fate of the above scar states and their signatures under perturbation. 
Moreover, we also want to examine if the $E = 0$ degenerate manifold has any relevance to the robustness of the scar states.
Thus, we examine the states $|\Gamma_{21} \ra$ and $|\Gamma_{I} \ra \equiv (|\Gamma_{11} \ra - |\Gamma_{22} \ra)/\sqrt{N}$, where $N = 2 - \frac{4}{3^{L/2}+(-1)^{L/2}}$ is the normalization factor (see Appendix~\ref{app:summaryexactscar}).
Both states have inversion quantum number $I = -(-1)^{L/2}$, and $|\Gamma_{I} \ra$ additionally has the particle-hole quantum number $\mathcal{C} = -(-1)^{L/2}$.
To compare their behavior against some thermal or chaotic state, we will also examine the state $|\Gamma_{\text{th}} \ra$, which is picked as the eigenstate with eigen-index three more than the index of $|\Gamma_{21} \ra$ in the $I = -(-1)^{L/2}$ symmetry sector.

The simplest inversion-symmetric perturbation that has the property $\hat{\mathcal{C}} \hV \hat{\mathcal{C}} = -\hV$ is
\begin{align} 
    \hV &= X_1 P_2 Z_3 + \sum_{j=2}^{L-2} P_{j-1} X_j P_{j+1} Z_{j+2} \notag \\
    &+ \sum_{j=3}^{L-1} Z_{j-2} P_{j-1} X_j P_{j+1} + Z_{L-2} P_{L-1} X_L ~.
    \label{eq:VPXPZ}
\end{align}
This perturbation was first studied in Ref.~\cite{khemaniSignatures2019}.
It was identified that at $\lambda \approx -0.02$, $\hH$ is close to some unknown integrable point.
Moreover, in the periodic boundary condition version of $\hH$, at $\lambda \approx -0.053$, a set of nearly perfect scar states was numerically found to lead to nearly perfect revivals  \cite{choiEmergent2019}\footnote{However, in the open boundary condition, at $\lambda = -0.053$, the revival dynamical signatures are less prominent compared to the periodic boundary condition case and the scar states are not strongly ``decoupled'' from the rest of the spectrum.}.

\section{Signatures of the scar states under perturbation}\label{sec:signatures}
\begin{figure*}
    \includegraphics[width=0.65\columnwidth]{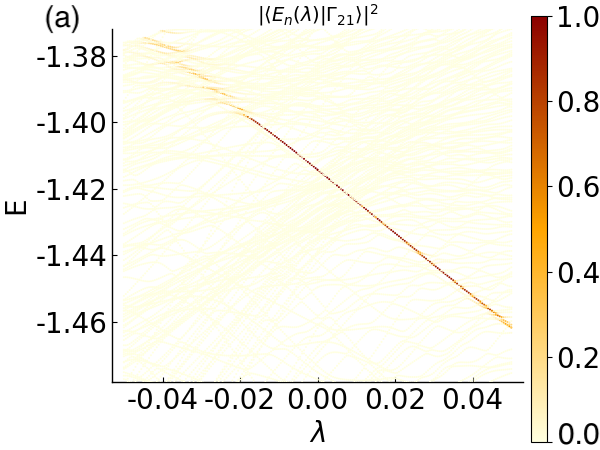}
    \includegraphics[width=0.65\columnwidth]{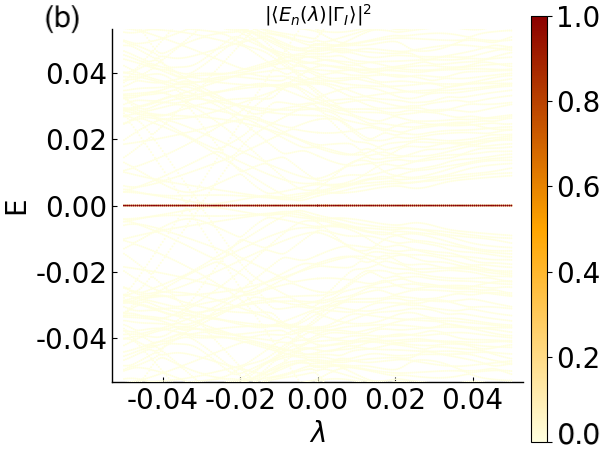}
    \includegraphics[width=0.65\columnwidth]{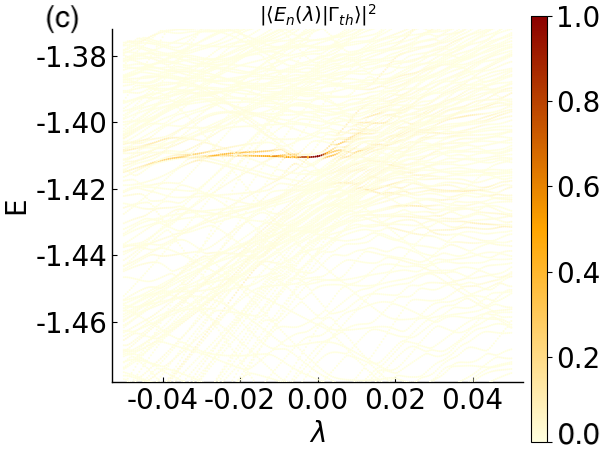}
    \caption{Intensity plot of the squared overlaps of the exact eigenstates of the perturbed Hamiltonian at size $L=20$ and symmetry sector $I=-1$ with: (a) and (b) the unperturbed scar states, $|\la E_n(\lambda)|\Gamma_{21} \ra|^2$ and $|\la E_n(\lambda)|\Gamma_{I} \ra |^2$ respectively, and (c) the unperturbed (presumably) thermal state, $|\la E_n(\lambda)|\Gamma_\text{th} \ra|^2$.
    The horizontal axis is the perturbation strength $\lambda$ while the vertical axis shows eigenstate energies.
    It appears that the scar states $|\Gamma_{21} \ra$ and $|\Gamma_I \ra$ hybridize with other states relatively weaker compared to the thermal state $|\Gamma_\text{th} \ra$, and exhibit some robustness in the finite-size ED spectrum.
    Note that for this system size when going from $\lambda = 0$ to $\lambda = 0.01$, the scar state $|\Gamma_{21} \ra$ crosses roughly $40$ states while still maintaining its fidelity.
    At $\lambda \approx -0.02$, the system is near some approximate integrability point.
    }
    \label{fig:ovespaghetti}
\end{figure*}

\begin{figure*}
    \includegraphics[width=\columnwidth]{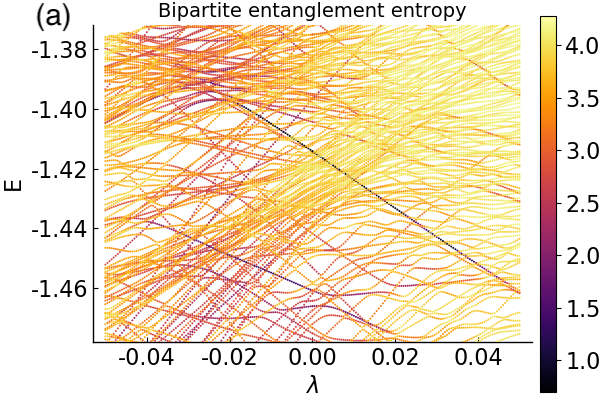}
    \includegraphics[width=\columnwidth]{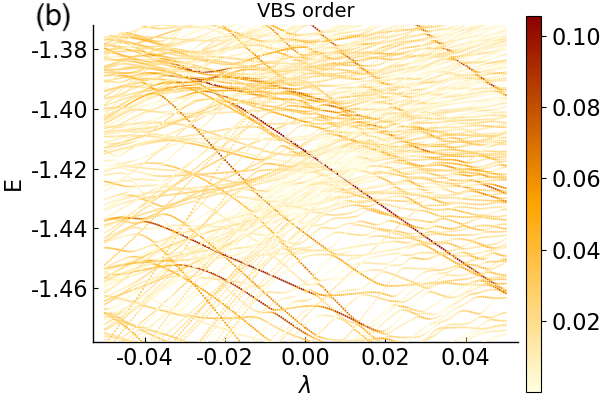}
    \caption{Apparent robustness of the scar signatures under the perturbation in the same system as in Fig.~\ref{fig:ovespaghetti}.
    The exact scar state $|\Gamma_{21} \ra$ has small (area-law) bipartite entanglement entropy and finite VBS order defined in Eq.~(\ref{eqn:MVBS}). 
    We therefore monitor these two signatures in panels (a) and (b) respectively.
    The descendant of the $|\Gamma_{21} \ra$ state is clearly visible in the range $-0.02 \lesssim \lambda \lesssim 0.05$---compare with Fig.~\ref{fig:ovespaghetti}(a).
    }
    \label{fig:EEspaghetti}
\end{figure*}

To see how robust the scar states are under the perturbation Eq.~(\ref{eq:VPXPZ}), we first examine the loss of the fidelity as we increase $\lambda$ in an open chain of length $L=20$.
In Fig.~\ref{fig:ovespaghetti} we examine the overlaps of exact eigenstates of the perturbed Hamiltonian with the unperturbed states, $|\la E_n(\lambda)|\Gamma \ra|^2$, where $|E_n(\lambda) \ra$ runs over eigenstates of $\hH = \hH_0 + \lambda \hV$, while $|\Gamma \ra$ is $|\Gamma_{21} \ra$, $|\Gamma_{I} \ra$, or $|\Gamma_{\text{th}} \ra$ in panels (a), (b), or (c) respectively.
In Fig.~\ref{fig:EEspaghetti}, we examine the bipartite entanglement entropy of $|E_n(\lambda) \ra$ and the VBS order parameter $\la E_n(\lambda)| \hat{M} |E_n(\lambda) \ra$ (associated with the unperturbed scar state $|\Gamma_{21} \ra$) measured in the exact eigenstates of the perturbed Hamiltonian as a function of the perturbation strength.

In Fig.~\ref{fig:ovespaghetti}(a), $|\la E_n(\lambda)|\Gamma_{21} \ra|^2$ clearly shows some apparent robustness under the perturbation [the numerical values can be seen in Fig.~\ref{fig:FidelityPert}(a)]. 
In the region $\lambda > 0$, there is a single perturbed eigenstate which can be traced back to $|\Gamma_{21} \ra$, despite multiple avoided level crossings when the perturbation strength is increased.
On the other hand, on the $\lambda < 0$ side, $|\Gamma_{21} \ra$ has significant overlap with several states near the approximate integrable point $\lambda \approx -0.02$~\cite{khemaniSignatures2019}. The spread in overlap makes it difficult to identify a single perturbed state associated with the unperturbed state $|\Gamma_{21} \ra$.
As a comparison, in Fig.~\ref{fig:ovespaghetti}(c), we show the fidelity loss of $|\Gamma_{\text{th}} \ra$. 
We can see that the overlaps of $|\Gamma_{\text{th}} \ra$ spread over multiple states at a smaller value of $\lambda$ as compared to those of $|\Gamma_{21} \ra$. This suggests that $|\Gamma_{\text{th}} \ra$ hybridizes more strongly with the other states in the spectrum upon perturbation.

In Fig.~\ref{fig:ovespaghetti}(b), $|\Gamma_{I} \ra$ appears to be even more robust under the perturbation than $|\Gamma_{21} \ra$.
Here, the overlaps between the $|E_n(\lambda)=0 \ra$ states and $|\Gamma_{I} \ra$ are summed up and displayed in the figure.
That is, we show the weight of $|\Gamma_{I} \ra$ projected into the perturbed $E=0$ manifold.
The numerical values can be seen more easily in Fig.~\ref{fig:FidelityPert}(b).

In Figs.~\ref{fig:EEspaghetti}(a) and \ref{fig:EEspaghetti}(b), we examine the bipartite entanglement entropy and the VBS order of the exact perturbed eigenstates near the unperturbed scar state $|\Gamma_{21} \ra$ under the perturbation.
Again, on the $\lambda > 0$ side, the perturbed eigenstates corresponding to the highest overlap with $|\Gamma_{21} \ra$ also shows low entanglement entropy and significant VBS order. 
Note that at $\lambda \approx -0.02$, the entire spectrum shows features of low entanglement, which is again a manifestation of the proximity to some integrable point~\cite{khemaniSignatures2019}.

\subsection{Perturbation theory}\label{subsec:perturb}
Here we show that the eigenstates which have high overlaps with the unperturbed states $|\Gamma_{21} \ra$ or $|\Gamma_{I} \ra$ are perturbed version of $|\Gamma_{21} \ra$ or $|\Gamma_{I} \ra$.
The fact that such a perturbation theory is controlled for our system sizes is a manifestation of the effective weakness of the perturbation as far as the scar states are concerned, which we will further quantify below.
However, we expect this perturbation theory to fail for large enough $L$ for any $\lambda \neq 0$ (see Sec.~\ref{subsec:finitesizescaling}).

The standard non-degenerate perturbation theory to $N$-th order gives the perturbative corrections to the unperturbed energy $E_\Gamma^{(0)}$ of the state $|\Gamma \ra$ as $E_{\text{pert.}}^{[N]} = E_\Gamma^{(0)} + \sum_{m=1}^N \lambda^m E_\Gamma^{(m)}$. 
Up to third order, we have
\begin{align}\label{eqn:perturbation_energy}
    E_\Gamma^{(1)} &= \la \Gamma| \hV |\Gamma \ra ~, \notag \\
    E_\Gamma^{(2)} &= \sum_{n \neq \Gamma} \frac{|\la n^{(0)}| \hV |\Gamma \ra|^2}{E_\Gamma^{(0)} - E_n^{(0)} } ~, \notag  \\
    E_\Gamma^{(3)} &= \sum_{n \neq \Gamma} \sum_{m \neq \Gamma} \frac{\la \Gamma| \hV |n^{(0)} \ra \la n^{(0)}| \hV |m^{(0)} \ra \la m^{(0)}| \hV |\Gamma \ra }{(E_\Gamma^{(0)} - E_n^{(0)}) (E_\Gamma^{(0)} - E_m^{(0)})} \notag \\
    &- \la \Gamma| \hV |\Gamma \ra \sum_{n \neq \Gamma} \frac{|\la n^{(0)}| \hV |\Gamma \ra|^2}{(E_\Gamma^{(0)} - E_n^{(0)})^2} ~.
\end{align}

\begin{figure}
    \includegraphics[width=\columnwidth]{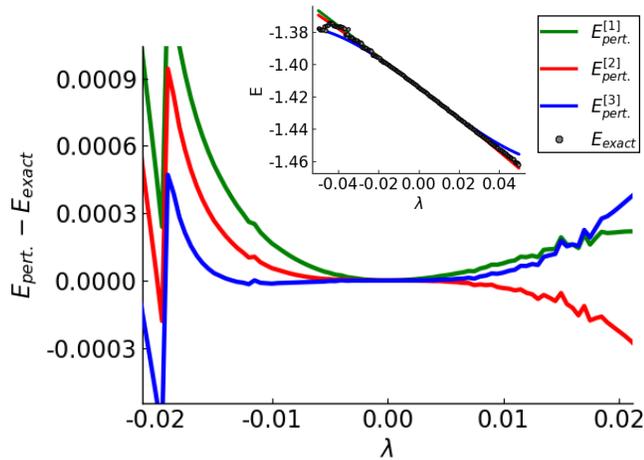}
    \caption{The energy difference between the perturbation theory $E_{\text{pert.}}$ (1st to 3rd order) and the ED result $E_{\text{exact}}$, where $E_{\text{exact}}$ is the energy of the eigenstate with the maximum overlap with $|\Gamma_{21} \rangle$.
    The perturbation theory gives accurate predictions of the energy of the perturbed eigenstate in the range of small perturbation strength $\lambda$. 
    Inset: The energies obtained from ED compared to the perturbation theory predictions up to 3rd order.
    The dominant shift in the energy is already captured by the 1st order, i.e., the ``diagonal" part of $V$, which also suggests relative weakness of the off-diagonal matrix elements.
    See text for the discussion of the accuracy of the perturbation theory and also Appendix~\ref{app:diag_pert_theory} for an improved treatment incorporating the diagonal part of $V$ in the unperturbed part.
    }
    \label{fig:EnergyPert}
\end{figure}

For $|\Gamma_{21} \ra$, we numerically calculate the perturbed energies $E_{\text{pert.}}^{[N]}$, $N=1,2$ and $3$ and compare them to the ED energy $E_{\text{exact}}$ as shown in Fig.~\ref{fig:EnergyPert}.
The ED energy $E_{\text{exact}}$ is the energy of the eigenstate that has the maximum overlap with $|\Gamma_{21} \ra$.
From the figure, we see that the perturbed result agrees very well with the ED result for small perturbation strength $\lambda$.
Note that as mentioned previously, the weight of $|\Gamma_{21} \ra$ is spread over several states on the $\lambda < 0$ side, making it hard to single out a particular eigenstate for the energy comparison, resulting in some non-smoothness in $E_{\text{exact}}$.

For $|\Gamma_{I} \ra$, we need to use degenerate perturbation theory since $|\Gamma_{I} \ra$ resides in the $E=0$ degenerate manifold.
However, if we choose the eigenbasis with definite inversion quantum number, then this basis is already appropriate for non-degenerate perturbation theory.
Indeed, an important corollary of the inversion symmetry of $\hV$ and of the property $\hat{\mathcal{C}} \hV \hat{\mathcal{C}} = -\hV$ is that $\la E_n=0| \hV |E_m=0 \ra = 0$ for any eigenstates $n$ and $m$ in the zero-energy manifold. 
The reason is that $\hV$ changes the particle-hole quantum number but preserves the inversion symmetry. 
As discussed in Sec.~\ref{sec:model}, the $E=0$ states have definite pairs of particle-hole and inversion quantum number.
This also implies that, to first order, the $E=0$ states are not hybridizing with each other.
Furthermore, using this property, we can prove that the perturbed wavefunction $|\Gamma_I^{(n)}\ra$ will have definite particle-hole quantum number $\mathcal{C}$ and hence $E_{\Gamma_I}^{(n)}=0$ to any order. 
We present the proof in Appendix.~\ref{app:EGammaIallorder}.
However, we emphasize that the proof does not imply the convergence of the perturbation theory in the thermodynamic limit.

Having established good agreement between the perturbed energy and the ED energy, we further show that the perturbed wavefunction is a good description for the ED wavefunction.
The (unnormalized) perturbed wavefunction to first order is given as 
\begin{equation}
    |\Gamma_{\text{pert.}} \ra = |\Gamma \ra + \lambda \sum_{n: E_n^{(0)} \neq E_\Gamma^{(0)}} |n^{(0)} \ra \frac{\la n^{(0)}| \hV | \Gamma \ra}{E_{\Gamma}^{(0)} - E_n^{(0)}} ~.
\end{equation}
We use this wavefunction to provide quantitative understanding of the fidelity loss obtained in ED.
Specifically, we calculate
\begin{align}\label{eqn:pert_fidelity}
    F_\text{pert.}(\lambda; \Gamma) &\equiv \frac{|\la \Gamma|\Gamma_{\text{pert.}} \ra|^2}{\la \Gamma_{\text{pert.}}|\Gamma_{\text{pert.}} \ra} \notag \\
    &= \left(1 + \lambda^2 \sum_{n: E_n^{(0)} \neq E_\Gamma^{(0)}} \frac{|\la n^{(0)}| \hV |\Gamma \ra|^2}{(E_n^{(0)} - E_{\Gamma}^{(0)})^2} \right)^{-1} ~,
\end{align}
and compare this with the ED result $F_\text{exact}(\lambda; \Gamma) \equiv |\la E_n(\lambda)|\Gamma \ra|^2$ (where the ED state is selected by maximizing the overlap).

\begin{figure}
    \includegraphics[width=\columnwidth]{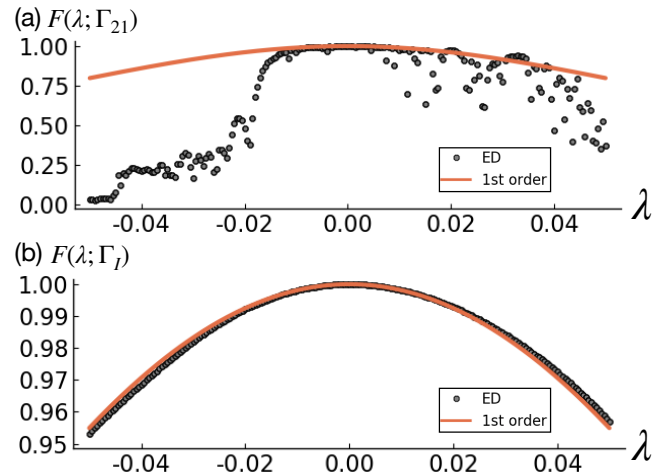}
    \caption{
    Comparison of the ED and the 1st-order perturbation theory results for the overlap between the descendant states and the unperturbed scar states, for (a) $|\Gamma_{21} \ra$ and (b) $|\Gamma_I \ra$ (in ED, the descendants are the eigenstates with the largest such overlaps) at $L=20$.
    The sudden drops in (a) on the $\lambda > 0$ side in the overlap in ED are caused by accidental hybridization that occurs very close to the avoided level crossings.
    On the $\lambda < 0$, the weight of $|\Gamma_{21} \ra$ is spread over several states, resulting in low overlap.
    The perturbation theory provides a good understanding for the eigenstate at small $\lambda$.
    }
    \label{fig:FidelityPert}
\end{figure}

In Figs.~\ref{fig:FidelityPert}(a) and (b), we compare the perturbation theory with the ED results, for $|\Gamma_{21} \ra$ and $|\Gamma_I \ra$ respectively.
In Fig.~\ref{fig:FidelityPert}(a), the ED result has some sudden jumps in the overlap on the $\lambda > 0$ side due to accidental events of (avoided) ``level crossings.''
When there is another eigenstate which happens to be very close in energy, the weight of $|\Gamma_{21} \ra$ will be spread over these states---known as accidental hybridization.
However, once the level crosses, the overlap recovers.
In fact, this phenomenon stems from the fact that $\la \Gamma_{21}| \hV |\Gamma_{21} \ra$ behaves differently from the nearby states $\la n^{(0)}| \hV |n^{(0)} \ra$.
This is also only possible when there are scar states present, since ETH would predict $\la n^{(0)}| \hV |n^{(0)} \ra$ to have the same value (up to some finite size correction) for $|n^{(0)} \ra$'s with the same energy density.
On the other hand, for $\lambda < 0$, the weight of $|\Gamma_{21} \ra$ is almost always spread over several states.
It is therefore less justified to single out a particular special state for the overlap comparison.

Turning to Fig.~\ref{fig:FidelityPert}(b),
since there is no level-crossing through the $E=0$ manifold, the ED result of the overlap to $|\Gamma_{I}\ra$ is smooth in $\lambda$.
We therefore see that, up to the accidental hybridization, the perturbation theory provides a good understanding for the perturbed ED scar states in this system size for both $|\Gamma_{21} \ra$ and $|\Gamma_{I} \ra$.

We note that technically, the Rayleigh-Schrodinger perturbation theory is not applicable when $\left|\frac{\la n^{(0)} | \hV | \Gamma \ra}{E_{n}^{(0)} - E_{\Gamma}^{(0)}} \right| \approx 1$. 
It is expected to already be not valid when there are (avoided) level crossings, as observed in the case of $|\Gamma_{21} \ra$. 
The good agreement between the eigenenergy $E_{\text{exact}}$ and the perturbation theory $E_{\text{pert.}}$ given the multiple level crossings is therefore indeed remarkable.
It reflects the smallness of the off-diagonal matrix elements while the crossing levels are ``pushed through" (as a function of $\lambda$) by the differing diagonal matrix elements.
To better explain the relative accuracy of the perturbation theory for this system size, we present a slightly modified version in Appendix~\ref{app:diag_pert_theory}.

\subsection{Finite-size scaling and eventual thermalization}
\label{subsec:finitesizescaling}

\begin{figure}
    \includegraphics[width=\columnwidth]{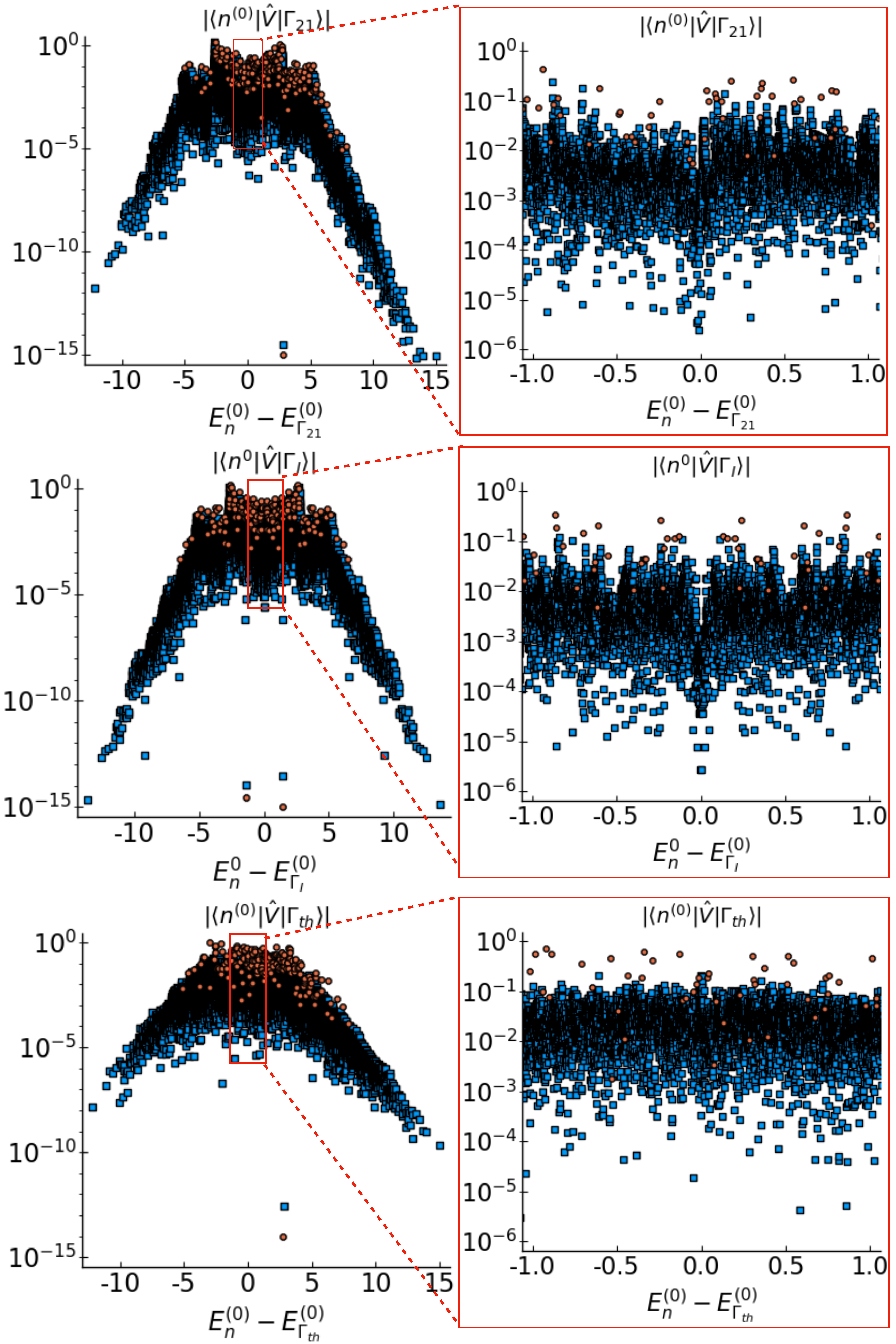}
    \caption{The distribution of the matrix elements $|\la n^{(0)}| \hV |\Gamma \ra|$, with $|\Gamma \ra$ chosen as $|\Gamma_{21} \ra$, $|\Gamma_I \ra$, and $|\Gamma_{\text{th}} \ra$, shown for system sizes $L=12$ and $L=22$.
    In all cases, there is a rapid falloff outside some energy window because of the locality of $\hH_0$ and $\hV$.
    In the case of the scar states, we see strong ``horns'' at $|E_n^{(0)} - E_\Gamma^{(0)}| \approx \pm 2.6$ (attributed to matrix elements to other scars in the spectrum) and also some suppression for small $|E_n^{(0)} - E_\Gamma^{(0)}|$, while in the thermal state case the distribution is more uniform.
    Nevertheless, the typical values of the matrix elements decrease with $L$ comparably for the scar and thermal states.
    In the right panels, we zoom in to show more detailed features of the distribution.
    }
    \label{fig:offdiag}
\end{figure}

Despite the success of the perturbation theory in describing $L=20$ ED results, here we study the finite-size scaling of the off-diagonal matrix elements and argue that the exact scar states will hybridize with the other ETH-satisfying states and thermalize eventually.
First, we examine the distribution of the amplitudes of the off-diagonal matrix elements $|\la n^{(0)}| \hV |\Gamma \ra|$ for each of the three states $|\Gamma \ra = |\Gamma_{21} \ra$, $|\Gamma_I \ra$, and $|\Gamma_\text{th}\ra$.
This is plotted on the left side in Fig.~\ref{fig:offdiag} with the zoomed-in scale shown on the right side.
(Since $\la E_n^{(0)}\!=\!0| \hV |\Gamma_I \ra = 0$, we omit these data.)
This figure gives us the idea about how the perturbation $V$ hybridizes the eigenstates of $H_0$, and how the overall magnitude of the matrix elements changes with the system size. 
(For clarity, only sizes $L=12$ and $L=22$ are shown.)

\begin{figure}
    \includegraphics[width=\columnwidth]{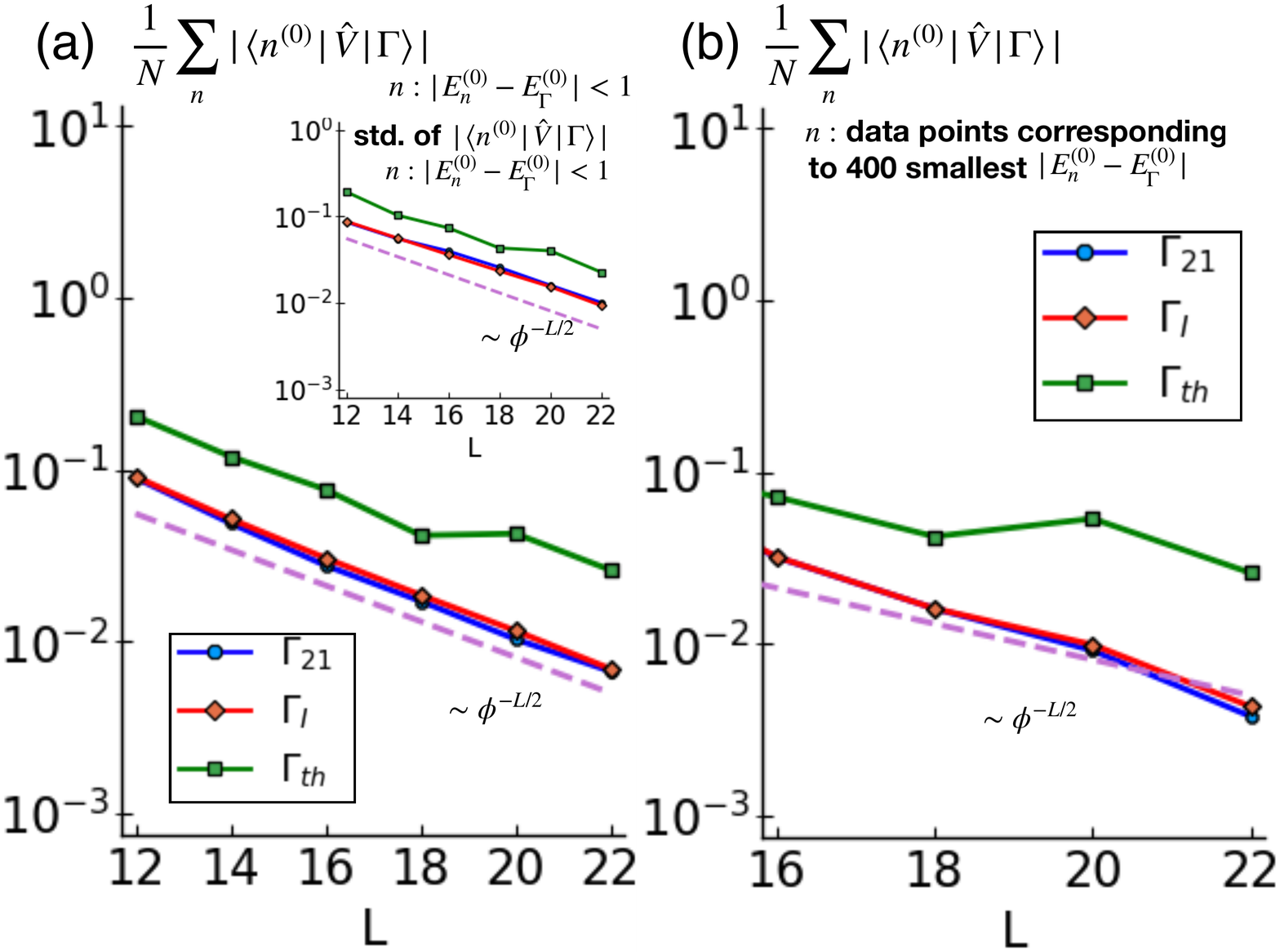}
    \caption{Finite-size scaling of the matrix elements $|\la n^{(0)}| \hV |\Gamma \ra|$ with $|\Gamma \ra$ being $|\Gamma_{21} \ra$, $|\Gamma_{I} \ra$, or $|\Gamma_{\text{th}} \ra$.
    (a) Average of $|\la n^{(0)}| \hV |\Gamma \ra|$ over $n$ such that the energy is within the window $|E_n^{(0)} - E_{\Gamma}^{(0)}| < 1$.
    Inset: the scaling of the standard deviation of $|\la n^{(0)}| V |\Gamma \ra|$ over $n$ within energy window $|E_n^{(0)} - E_{\Gamma}^{(0)}| < 1$.
    (b) Average of $|\la n^{(0)}| \hV |\Gamma \ra|$ over $n$ such that the energy difference $|E_n^{(0)} - E_{\Gamma}^{(0)}|$ is among the smallest $400$.
    For all states---thermal or scar---the behavior of the off-diagonal matrix elements satisfy the scaling predicted by the random matrix theory.
    }
    \label{fig:offdiagscaling}
\end{figure}

In the left panels, we see that for $|\la n^{(0)}| \hV |\Gamma_{21} \ra|$ and $\la n^{(0)}| \hV |\Gamma_{\text{th}} \ra$, there is a state in each case with zero overlap (showing up at numerical error threshold $\sim 10^{-14}$) with $\hV |\Gamma_{21} \ra$ or $\hV |\Gamma_{\text{th}} \ra$ respectively.
These states are just $\hat{\mathcal{C}}|\Gamma_{21} \ra$ and $\hat{\mathcal{C}}|\Gamma_{\text{\text{th}}} \ra$ respectively.
One can easily see that $\la \Gamma| \hat{\mathcal{C}} \hV |\Gamma \ra = -\la \Gamma| \hV \hat{\mathcal{C}} |\Gamma \ra = -\la \Gamma| \hat{\mathcal{C}} \hV |\Gamma \ra ^{*} = 0$, where in the last equality we used time-reversal-like symmetry, or the fact that $\la \Gamma| \hV \hat{\mathcal{C}} |\Gamma \ra$ is a real number.
On the other hand, for $|\Gamma_{I} \ra$, we see two states with zero matrix elements $\la n^{(0)}| \hat{V} |\Gamma_{I} \ra$.
These states are in fact $|\Gamma_{21} \ra$ and $|\Gamma_{12} \ra$.
However, these matrix elements are zero due to the special structure of the states, instead of some symmetry reasoning.
Also note that this is not reflected in the plot of $|\la n^{(0)}| \hat{V} |\Gamma_{21} \ra|$, because the $|E_n^{(0)}=0 \ra$ states are obtained through numerical diagonalization, and $|\Gamma_{I} \ra$ is a superposition of these states. 

For $|\Gamma_{\text{th}} \ra$, we see that the distribution of $|\la n^{(0)}| \hV |\Gamma_{\text{th}} \ra|$ is fairly uniform and does not have any features within the energy window $|E_n^{(0)} - E_{\Gamma_{\text{th}}}^{(0)}| \lesssim 1$.
On the other hand, we see that for $|\Gamma_{21} \ra$ and $|\Gamma_I \ra$, there is some ``horn" structure at $|E_n^{(0)} - E_\Gamma^{(0)}| \approx 2.6$.
This is due to the strong connection between the exact scars and other quasiparticle-like excitation states on top of the exact scar states. 
(See Ref.~\cite{linExact2019,suraceLattice} for such a ``quasiparticle picture'' of the other nonexact scar states in the PXP model in the spirit of single mode approximation and its multi-mode generalization.)
In all three cases, there is a rapid dropoff of the matrix elements for $|E_n^{(0)} - E_{\Gamma_{\text{th}}}^{(0)}| \gtrsim 5$, which reflects the locality of the  Hamiltonian--- see Appendix~\ref{app:distr_matr_elem} for details.

We can see some suppression in the amplitudes of the matrix elements $|\la n^{(0)}| \hV |\Gamma_{21} \ra|$ and $|\la n^{(0)}| \hV |\Gamma_{I} \ra|$ near zero $|E_n^{(0)} - E_\Gamma^{(0)}|$.
We think that for these system sizes, this suppression in the off-diagonal matrix element amplitudes further assists the validity of the perturbation theory description in Fig.~\ref{fig:FidelityPert} discussed earlier.
However, by examining the distributions in more narrow energy windows, we find that this only affects the connection to the thermal states quantitatively and not qualitatively.

We further use the statistics of $|\la n^{(0)}| \hV |\Gamma \ra|$ to argue for the eventual thermalization of the original unperturbed eigenstates.
For an ETH system under a perturbation $\hV$, within some energy window $|E_n^{(0)} - E_m^{(0)}| < C$, one expects the off-diagonal matrix elements $\la n^{(0)}| \hV |m^{(0)} \ra$ to scale as ${\cal D}_L^{-1/2}$, where ${\cal D}_L$ is the dimension of the Hilbert space for the chain of length $L$ (${\cal D}_L \sim \phi^L$ for the Rydberg-blockaded chain).
The reason is that if $|n^{(0)} \ra$ is thermal or chaotic, than $|n^{(0)} \ra$ behaves essentially like a random vector in the Hilbert space when considering off-diagonal matrix elements.
We can view $\hat{V} |m^{(0)} \ra$ as some fixed ``direction" in the Hilbert space, and the overlap between a random vector and a vector in some direction will be of order ${\cal D}_L^{-1/2}$.
Note that we needed only the state $|n^{(0)} \ra$ to be ``thermal'' (i.e., essentially ``random'') while the other state $|m^{(0)} \ra$ can be either thermal or non-thermal as long as $\hat{V} |m^{(0)} \ra$ is not somehow special, e.g., $|m^{(0)} \ra$ does not happen to be an exact eigenstate of $\hat{V}$.
On the other hand, the density of states at energy density corresponding to infinite temperature grows as $\sim{\cal D}_L$.
Therefore the ratio between the typical off-diagonal matrix elements and the level-spacing will grow as ${\cal D}_L^{1/2}$, which signals the strong hybridization or thermalization.

A well known counterexample which circumvents thermalization is many-body localization \cite{PhysRevB.75.155111,serbynLocal2013,husePhenomenology2014,chandranConstructing2015,nandkishoreManybody2015,aletManybody2018,chenHow2018,RevModPhys.91.021001}.
In such systems, the off-diagonal matrix element for nearby states scales as $e^{-L/\xi}$, where $\xi$ is some localization length.
In a crude estimate, if $\xi < L/\ln {\cal D}_L$, the ratio between typical off-diagonal matrix element and level-spacing will in fact decrease with $L$ \cite{serbynCriterion2015a}. 
The eigenstates of many-body-localized systems are thus perturbatively accessible at any $L$.

Returning to our setting of following the fate of the scar states under the perturbation, we have already observed that they generically have nonzero matrix elements to all eigenstates with the same quantum numbers, and that they appear to lose fidelity upon increasing the perturbation strength or the system size.
We also expect that the scar states are exceptions in the unperturbed PXP model while most of the eigenstates are ETH-satisfying. 
The above scaling of the matrix element between a scar state and an ETH state with $L$ should therefore hold.
Nevertheless, we would like to check this explicitly and examine more detailed quantitative features of such connections.
Accordingly, we next study the finite-size scaling of the off-diagonal matrix elements.

In Figs.~\ref{fig:offdiagscaling}(a) and (b), we show the statistics of the off-diagonal matrix elements. 
In Fig.~\ref{fig:offdiagscaling}(a), we averaged $|\la n^{(0)}| \hV |\Gamma \ra|$ in the energy window $|E_n^{(0)} - E_\Gamma^{(0)}| < 1$, but excluding $|E_n^{(0)} = 0 \ra$ states in the $|\Gamma_{I} \ra$ case.
Not surprisingly, $|\la n^{(0)}| \hV |\Gamma_{\text{th}} \ra|$ shows the $\phi^{-L/2}$ scaling predicted by the random matrix theory description.
Importantly, we also see the same scaling for the scar states $|\la n^{(0)}| \hV |\Gamma_{21} \ra|$ and $|\la n^{(0)}| \hV |\Gamma_{I} \ra|$, but with smaller amplitude than for the thermal state.
To avoid the arbitrariness of specifying the energy window for averaging, in Fig.~\ref{fig:offdiag}(b), we take 400 matrix elements corresponding to 400 states with the smallest $|E_n^{(0)} - E_{\Gamma}^{(0)}|$. 
(We do not show $L=12$ and $L=14$ data because $400$ states is larger than the Hilbert space dimension for $L=12$ and larger than half of the Hilbert space dimension for $L=14$.)
For both averaging protocols, we see that the off-diagonal matrix element scaling of the thermal $|\Gamma_{\text{th}} \ra$ and scar states $|\Gamma_{21}\ra$ and $|\Gamma_{I}\ra$ all satisfy the random matrix theory prediction, while the overall amplitudes for the scar states are smaller.
This also explains the apparent weaker hybridization observed in Fig.~\ref{fig:ovespaghetti}:
the apparent robustness of the scars compared to the thermal state is due to the smaller amplitude of the off-diagonal matrix elements.

The relevant quantitative difference is even stronger than suggested by the averages in Fig.~\ref{fig:offdiagscaling} because the number of states being averaged covers a wider window than the ``dip" close to zero $|E_n^{(0)} - E_\Gamma^{(0)}|$ seen in Fig.~\ref{fig:offdiag}, while the states inside this dip play a larger role in the perturbed wavefunction at these sizes.
As mentioned earlier, looking more closely at the matrix elements inside the dip (e.g., averaging over smaller number of states, all the way down to just few closest states), we do not see any faster decay with $L$ than expected in the random matrix theory.

To conclude, from the above scaling argument comparing the off-diagonal matrix elements and the level-spacing, we expect the scar states will eventually thermalize in the thermodynamic limit.
In fact, in Appendix~\ref{app:spin-1XY}, we perform a similar analysis in the spin-1 XY model~\cite{schecterWeak2019}, and find similar finite-size scaling behavior.
Finally, we also note that comparison between the $|\Gamma_{21} \ra$ and $|\Gamma_I \ra$ cases suggests that the degenerate $E=0$ manifold (to which the latter state belongs) is in fact irrelevant to the hybridization of the scar states.

\section{Slow thermalization of local observables}\label{sec:prethermalization}
Despite the eventual thermalization of the scar states under perturbations, in the following, we show numerically and analytically, that the signatures of the scars can survive up to some long time set by the perturbation strength even in the thermodynamic limit.
This slow thermalization is analogous to that in systems that are weakly perturbed from integrability \cite{bertiniPrethermalization2015,bertiniThermalization2016} or that have weak breaking of some conservation laws \cite{mallayyaPrethermalization2019,PhysRevX.8.021030}.
Specifically, we consider the following general global quench setting:
we consider the initial state as an exact scar state $|\Gamma \ra$ of some Hamiltonian $\hH_0$, and let it evolve under the perturbed Hamiltonian $\hH = \hH_0 + \lambda \hV$.
Since $|\Gamma \ra$ is a scar state, there is some local observable $\hat{m}$ which is nonthermal.
We then examine how quickly such an observable behaves at late times, and if it reaches its thermal value.

\subsection{Slow decay of VBS order in numerics}\label{subsec:tebd}
\begin{figure*}
    \centering
    \includegraphics[width=0.7\columnwidth]{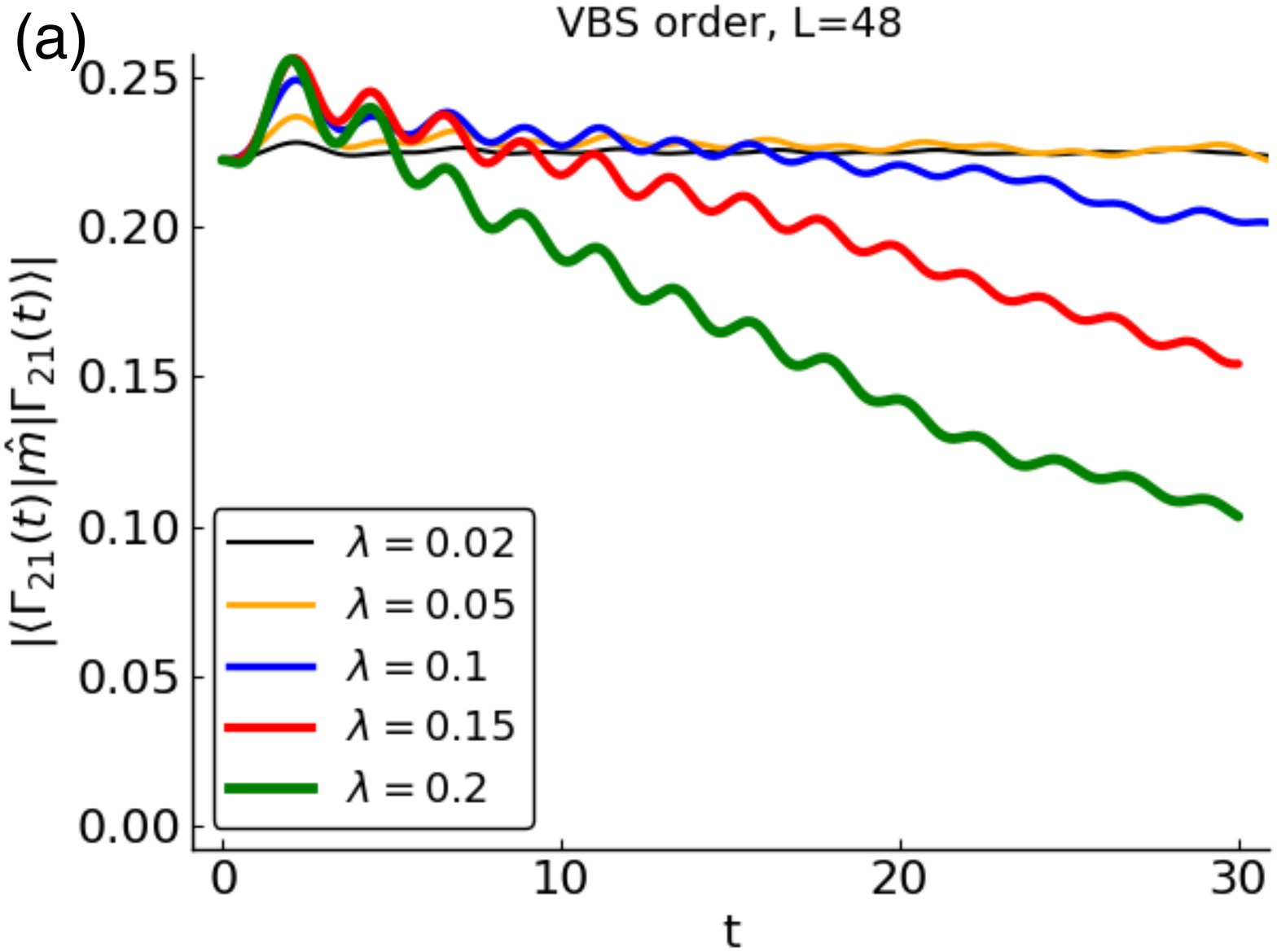}
    \includegraphics[width=0.7\columnwidth]{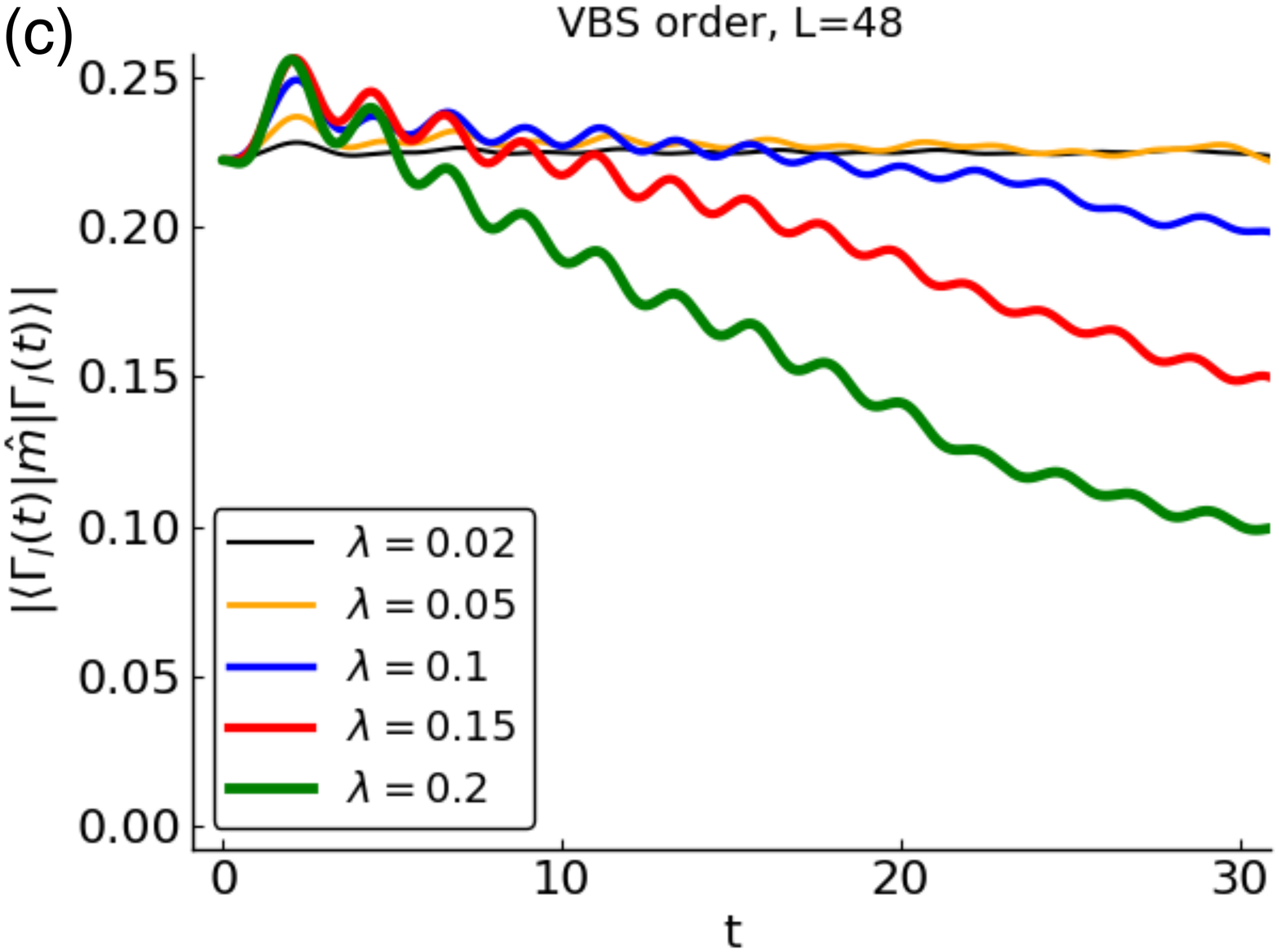}
    \includegraphics[width=0.6\columnwidth,height=4.5cm]{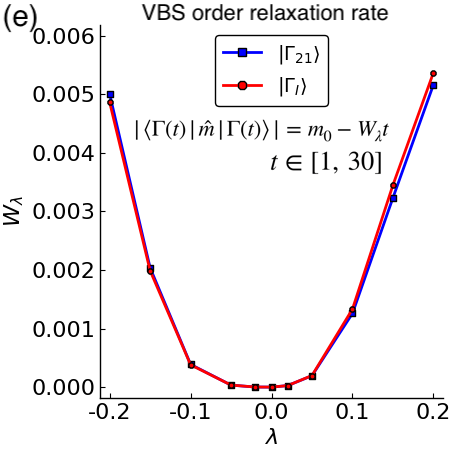}
    \includegraphics[width=0.7\columnwidth]{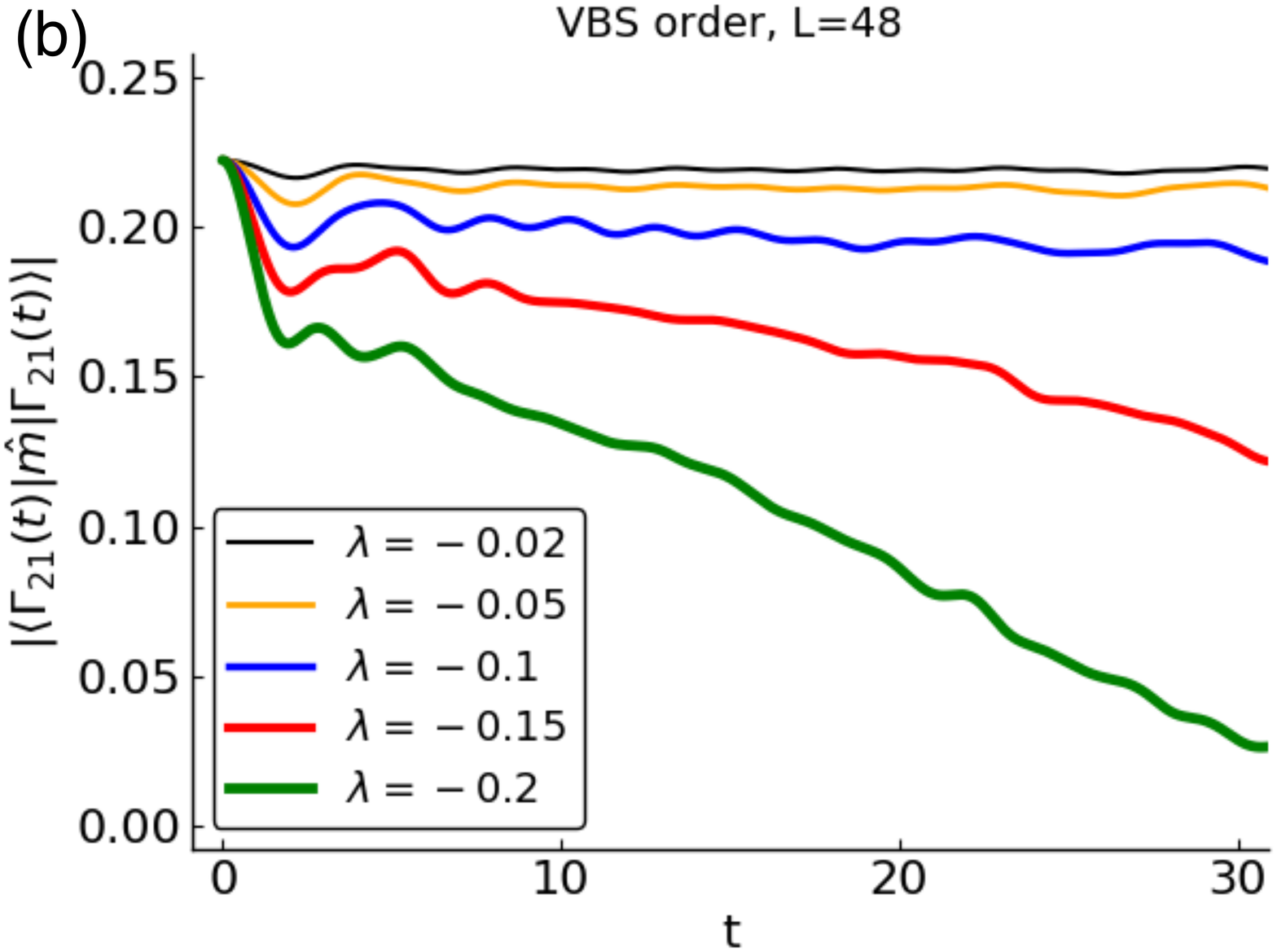}
    \includegraphics[width=0.7\columnwidth]{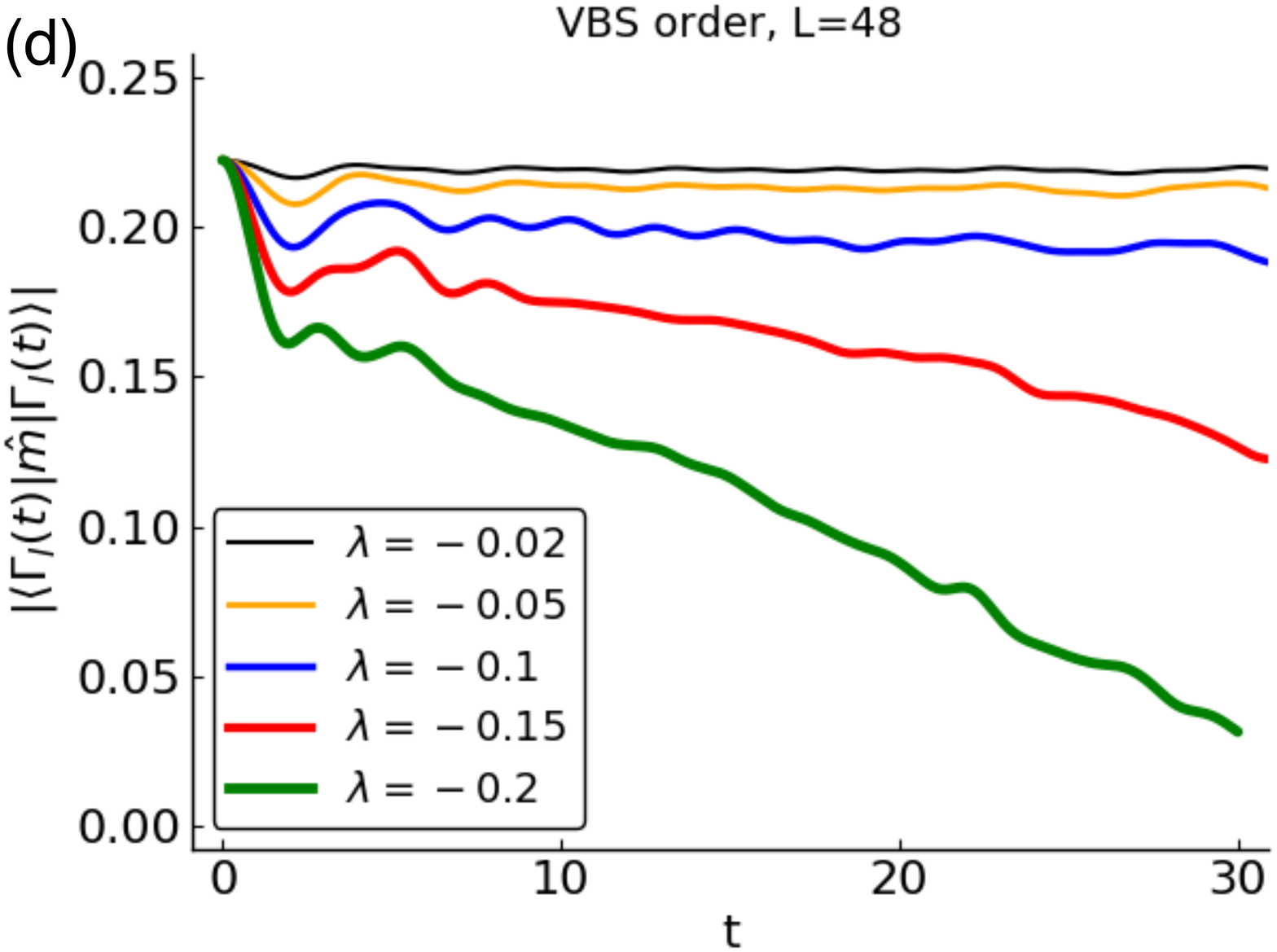}
    \includegraphics[width=0.6\columnwidth,height=4.5cm]{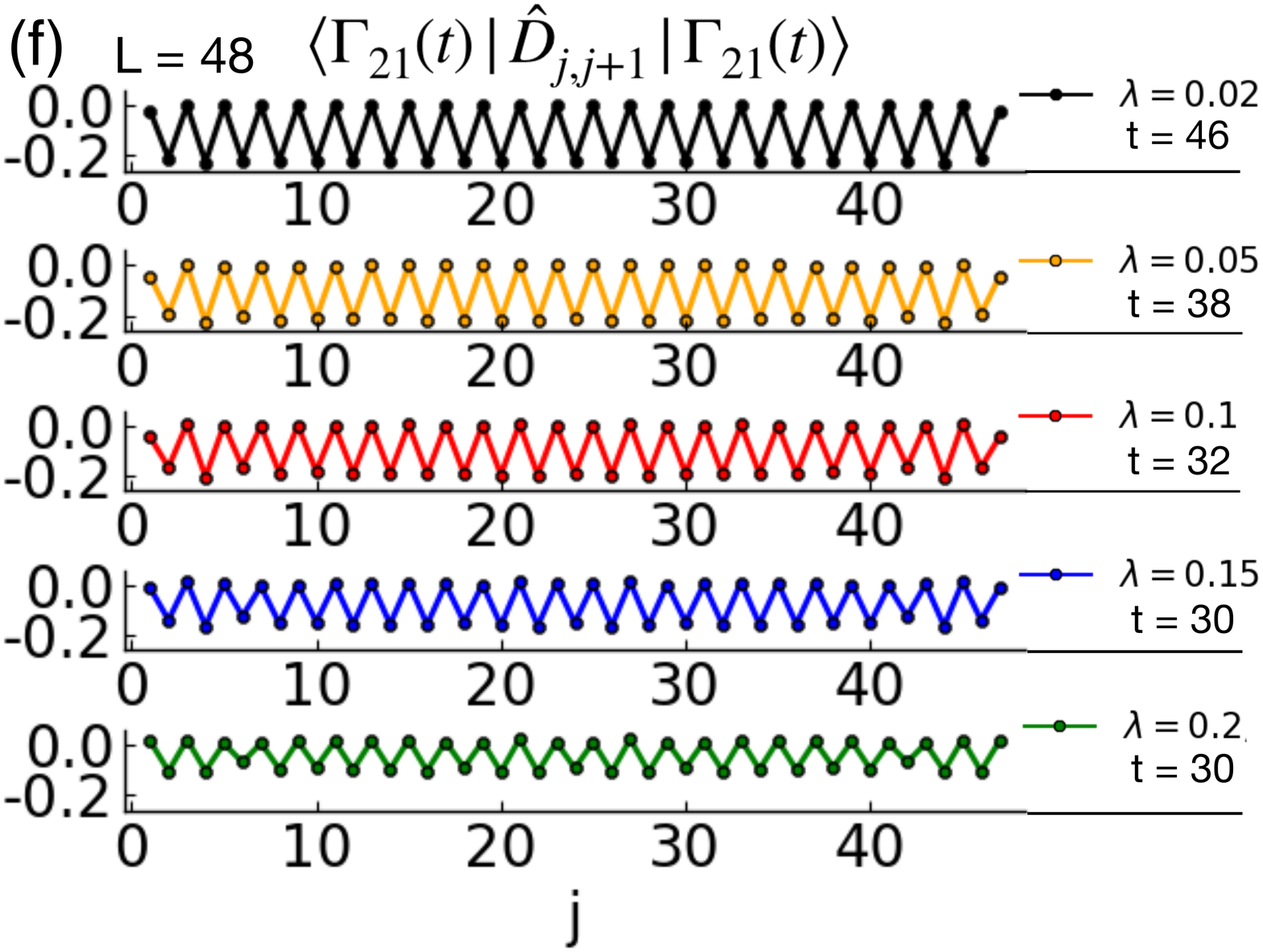}
    \caption{Relaxation of the VBS order parameter measured in the middle of the chain, $\hat{m} = \hat{D}_{L/2,L/2+1} - \hat{D}_{L/2+1,L/2+2}$, for system size $L=48$.
    (a) and (b): Initial state $|\Gamma_{21} \ra$ with parameters $\lambda > 0$ and $\lambda < 0$ respectively.
    (c) and (d): Initial state $|\Gamma_I \ra$ with parameters $\lambda > 0$ and $\lambda < 0$ respectively.
    From the dynamical data, there is no qualitative difference in the time evolution of the observable $\hat{m}(t)$ between the initial states $|\Gamma_{21} \ra$ and $|\Gamma_I \ra$. 
    We therefore conclude that the $E=0$ degenerate manifold 
    does not influence qualitatively the thermalization behavior of the scar state $|\Gamma_I \ra$ from this manifold.
    (e) We empirically estimate the relaxation rate $W_\lambda$ of the VBS order by fitting $|\la \Gamma(t)| \hat{m} |\Gamma(t) \ra| = m_0 - W_\lambda t$ using data in the time interval $t \in [1,30]$ and plot $W_\lambda$ vs the perturbation strength $\lambda$.
    (f) Snapshots of the dimer strength pattern $\la \Gamma_{21}(t)| \hat{D}_{j,j+1} |\Gamma_{21}(t) \ra$ for several $\lambda > 0$, taken at the largest time (or even larger) shown in panel (a). 
    These show that measuring the VBS amplitude in the middle, $\hat{m}$, is representative of the VBS order across the chain.
    }
    \label{fig:tebd}
\end{figure*}

First, we numerically show the slow thermalization using time-evolved block decimation (TEBD) method \cite{vidalEfficient2003} \footnote{In all of our TEBD calculation, we use Trotter step $\Delta t =0.02$ and bond-dimension $\chi =500$.}.
We choose the initial state as $|\Gamma_{21} \ra$ or $|\Gamma_I \ra$ and evolve it under $\hH = \hH_0 + \lambda \hV$, where $\hH_0$ and $\hV$ are given in Eqs.~(\ref{eqn:PXP}) and~(\ref{eq:VPXPZ}) respectively.
We measure the VBS order parameter in the middle of the system,
\begin{align}
\hat{m} = \hat{D}_{L/2, L/2+1} - \hat{D}_{L/2+1, L/2+2} ~.
\end{align}
Figures~\ref{fig:tebd} (a) to (d) show the time evolution of this observable for the two initial states and different perturbation strengths $\lambda$.

At short times, the VBS order deviates from the initial value as $t^2$.
This can be easily understood from the time-reversal-like symmetry.
Considering the Taylor series
\begin{equation}
    \la \Gamma (t)| \hat{m} |\Gamma(t) \ra = \sum_{n=0}^{\infty} \frac{1}{n!} \left[ \frac{d^{n}}{dt^n} \la \Gamma (t)| \hat{m} |\Gamma(t) \ra \Big|_{t=0} \right] t^{n} ~,
\end{equation}
the coefficient of $t^{1}$ is given by 
\begin{align}
\frac{d}{dt} \la \Gamma(t)| \hat{m} |\Gamma(t) \ra \Big|_{t=0} &= i \la \Gamma| [\hH, \hat{m}] |\Gamma \ra ~.
\end{align}
Now, $\la \Gamma(t)| \hat{m} |\Gamma(t) \ra$ is a real number by the hermiticity of the observable $\hat{m}$, while $\la \Gamma| [\hH, \hat{m}] |\Gamma \ra$ is a real number because the operators $\hat{H}$ and $\hat{m}$ have real-valued matrix elements in the Rydberg atom basis, and the wavefunction $|\Gamma \ra$ has real-valued amplitudes in this basis.
Hence, we conclude that $\frac{d}{dt} \la \Gamma(t)| \hat{m} |\Gamma(t) \ra|_{t=0} = 0$.
The coefficient of the $t^2$ growth is
\begin{equation}
\frac{d^2}{dt^2} \la \Gamma(t)| \hat{m} |\Gamma(t) \ra \Big|_{t=0} = -\la \Gamma| [\hH, [\hH, \hat{m}]] |\Gamma \ra ~;
\end{equation}
this is generically not zero, and its sign determines if $\la \Gamma(t) |\hat{m} |\Gamma(t) \ra$ curves up or down initially.

The initial $t^2$ behavior stops at around $t \approx 1$, and beyond this time the VBS order appears to relax almost linearly.
(The oscillations that are clearly visible on top of the overall decay are roughly at frequency $\omega \approx 3$ and can be traced to the ``horn" features in the distribution of the matrix elements discussed in Fig.~\ref{fig:offdiag}.)
At small $\lambda = 0.02$, the VBS order has an almost negligible relaxation rate.
As $\lambda$ increases, the relaxation of the VBS order becomes visible within the time window shown in the figure.
Note that in Figs.~\ref{fig:ovespaghetti} and \ref{fig:EEspaghetti}, we study the perturbation strength $|\lambda|\leq 0.05$.
Here we also study larger perturbation strengths to have noticeable VBS order relaxation.

Given the almost linear relaxation behavior of the VBS order, we empirically extract the relaxation rate by fitting the data via the least-squares method to 
\begin{equation}
    \la \Gamma(t)| \hat{m} |\Gamma(t) \ra = m_0 - W_\lambda t ~,    
\end{equation} 
in the time interval $t \in [1,30]$. Fig.~\ref{fig:tebd}~(e) plots the relaxation rate $W_{\lambda}$.
The relaxation rate appears to vanish faster than linearly at small $\lambda$, which corresponds to a relaxation time that diverges faster than $\lambda^{-1}$.
However, since the slope obtained from the linear fit at the smallest $|\lambda| \leq 0.05$ are numerically very small and may not be very reliable, we cannot extract the precise functional form.

In Fig.~\ref{fig:tebd}~(f), we show the snapshots of the dimer strength pattern $\la \Gamma_{21}(t)|\hat{D}_{j,j+1}|\Gamma_{21}(t)\ra$.
It clearly shows that the dimer strength pattern is uniform on even/odd sites, so $\hat{m}$ is indeed representative of the VBS order.

\begin{figure}
    \includegraphics[width=\columnwidth]{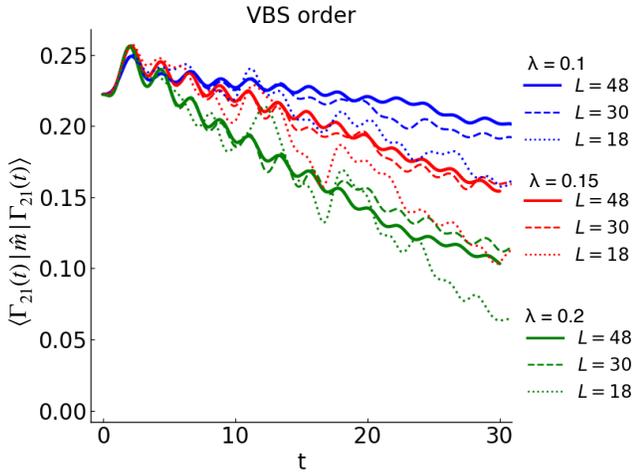}
    \caption{The decay of the VBS order $\hat{m} = \hat{D}_{L/2,L/2+1} - \hat{D}_{L/2+1,L/2+2}$ with time for several perturbation strengths and different system sizes.
    The data for sizes $L=30$ and $L=48$ are calculated by TEBD method with bond dimension $\chi=500$ and Trotter step $dt = 0.02$.
    The data for $L=18$ is calculated via ED.
    }
    \label{fig:VBSdifferentsizes}
\end{figure}

To understand the effects of finite size, in Fig.~\ref{fig:VBSdifferentsizes}, we show the VBS order decay $\la \Gamma_{21}(t)| \hat{m} |\Gamma_{21}(t) \ra$ for different system sizes. 
We can see that the traces of the small system sizes $L=18$ and $L=30$ are converging to the trace of $L=48$ up to the time of about $10$ or slightly larger.
We think that this time is what is often referred to as ``recurrence" time in ED studies of quantum many-body dynamics, which is roughly the time it takes for information to propagate across the entire system~\cite{banulsStrong2011,linQuasiparticle2017}.
In this picture, the $L=48$ data up to such time $t \simeq 10$ is already representative of the thermodynamic limit.
We therefore think that the almost linear relaxation behavior observed up to this time is already representative of the thermodynamic limit.
The relaxation is clearly visible for $\lambda \geq 0.1$, and we expect eventual thermalization for such perturbation strengths; the relaxation is not visible for the smaller $\lambda = 0.02$ and $0.05$.
Although we do not expect any qualitative changes as we vary $\lambda$, the characteristic relaxation time scale can grow as $\lambda$ approaches zero.
We therefore argue that our exact scar states eventually thermalize for any nonzero $\lambda$.

It is interesting to note that all sizes in Fig.~\ref{fig:VBSdifferentsizes} show qualitatively similar relaxation behavior that continues well beyond the above estimated recurrence times for these sizes.
(For the largest perturbation strength $|\lambda|=0.2$, the systematic relaxation for the smallest size $L=18$ stops around $t \simeq 50$, beyond which time the trace wanders non-systematically.)
At present, we do not have a good understanding of this observation. Nevertheless, it suggests that the presented time range in Figs.~\ref{fig:tebd} and \ref{fig:VBSdifferentsizes} may be representative of the thermodynamic limit behavior beyond the recurrence time $t \simeq 10$.

Finally, we note that the dynamical signature of the VBS order is essentially the same starting from either $|\Gamma_{21} \ra$ or $|\Gamma_I \ra$ for large enough system sizes. 
This is therefore another piece of evidence suggesting that the $E=0$ manifold does not have significant effects on the observable dynamical signatures of the scar states.

\subsection{Rigorous bound on the thermalization time}\label{subsec:rigorousbound}
We have shown numerically that small perturbation strength indeed gives slow thermalization of the VBS order.
However, as in most numerical calculations, the results may be strongly affected by finite-size effects, and one may worry that the slow thermalization may not be a true phenomenon in the thermodynamic limit.
Remarkably, in the following, we give a rigorous lower bound on the thermalization time for the scar states, even in the thermodynamic limit, that diverges when the perturbation strength goes to zero.
The following theorem in fact is valid in any dimension, though we will specialize it to one dimension first.

Recall that we consider the global quench setting as
\begin{equation}
\hH = \hH_0 + \lambda \hV ~, \quad 
|\Gamma(t) \ra = e^{-i \hH t} |\Gamma \ra ~.
\end{equation}
We also further assume that both $\hH_0$ and $\hV$ are local Hamiltonians, i.e., lattice sums of local terms, e.g., $\hV = \sum_j \hat{v}_j$ where $\hat{v}_j$ acts only on degrees of freedom near site $j$.

Consider a local observable $\hat{m}$, which we assume to be localized near the origin $j = 0$ (at the middle of the chain), and its expectation value, $\la \Gamma(t)| \hat{m} |\Gamma(t) \ra$.
Consider the time derivative
\begin{align}
\frac{d}{dt} \la \Gamma(t)| \hat{m} |\Gamma(t) \ra & = i \la \Gamma(t)| [\hH, \hat{m}] |\Gamma(t) \ra \\
& = i \la \Gamma| [\hH, e^{i \hH t} \hat{m} e^{-i \hH t}] |\Gamma \ra \\
& = i \la \Gamma| [\lambda \hV, e^{i \hH t} \hat{m} e^{-i \hH t}] |\Gamma \ra ~,
\label{eq:derivm}
\end{align}
where in the last line we used that $|\Gamma \ra$ is an eigenstate of $\hH_0$.

For $t = 0$, the right hand side is given by the expectation value of $[\lambda \hV, \hat{m}]$ in the initial state $|\Gamma \ra$.
Since this is a local operator, the expectation value is $\lambda$ times an $O(1)$ number, and the initial rate of change of the observable is thus of order $\lambda$.
(In the specific quench setting considered in the previous subsection, this term is actually zero because of the time reversal invariance of $\hH$, but we will not use this in the developments below.)
The small parameter $\lambda$ is present as a factor in Eq.~(\ref{eq:derivm}) at any time $t$, but we have to consider the possibility that it multiplies a function of $t$ that grows with time.

A conservative bound can be obtained by noting that the time-evolved operator $e^{i \hH t} \hat{m} e^{-i \hH t}$ is significantly spread only over region $|j| \leq v_{LR}\, t$, where $v_{LR}$ is the Lieb-Robinson velocity for the Hamiltonian $H$ \cite{lieb1972,hastingsSpectral2006,hastingsLocality}.
Therefore, this time-evolved operator can have a significant commutator only with local terms in $V$ that are inside this region, i.e., $\hat{v}_j$ with $|j| \leq v_{LR}\, t$.
Now, for any $j$, the norm of the operator $[\lambda \hat{v}_j, e^{i \hH t} \hat{m} e^{-i \hH t}]$ is bounded by $2 \|\lambda \hat{v}_j\| \|\hat{m}\|$, which is $\lambda$ times an $O(1)$ number.
Hence, the total contribution to Eq.~(\ref{eq:derivm}) from $|j| \leq v_{LR}\, t$ is bounded by $\lambda (c_0' + c_1 t)$, where $c_0'$ and $c_1$ are fixed $O(1)$ numbers (and for concreteness we take the spatial dimension to be $d = 1$).
On the other hand, using the Lieb-Robinson bounds $\|[\hat{v}_j, \hat{m}(t)]\| \leq c \exp[-a(j - v_{LR}\, t)]$, the total contribution from $|j| > v_{LR}\, t$ can be bounded by $\lambda c_0''$, where $c_0''$ is also a fixed $O(1)$ number.
Thus, we have
\begin{align}
 \left| \la \Gamma| [\lambda \hV, e^{i \hH t} \hat{m} e^{-i \hH t}] |\Gamma \ra \right| \leq \lambda (c_0 + c_1 t) ~. \label{eq:derivbound}
\end{align}

A minor point:
The Lieb-Robinson bounds are used here with respect to the full Hamiltonian $\hH = \hH_0 + \lambda \hV$, which depends on the parameter $\lambda$.
However, typically used estimates---effectively, upper bounds---of the Lieb-Robinson velocity use operator norms of local terms in the Hamiltonian and will depend smoothly on $\lambda$, which will only introduce a smooth $\lambda$ dependence in the parameters $c_0$ and $c_1$.
Hence, the leading $\lambda$ dependence of the above bound is essentially unchanged.
One can provide a slightly different argument using the Duhamel formula\cite{abaninRigorous2017} where this minor point does not arise at all.

The analysis so far is completely general and applies to any $\hH_0$ and any initial eigenstate $|\Gamma \ra$.
Let us now see how we can use it to lower-bound the persistence time of nonthermal properties when $|\Gamma \ra$ is a scar eigenstate of $\hH_0$.
In this case, $\la \Gamma| \hat{m} |\Gamma \ra$ differs by a finite amount from the expectation value of $\hat{m}$ in the nearby ``thermal'' eigenstates, which we will denote as $m_{\text{th.};\, \hH_0,\, E_0 = \la \Gamma| \hH_0 |\Gamma \ra}$.
We assume that $|\Gamma(t) \ra$ will eventually ``thermalize'' with respect to $\hH = \hH_0 + \lambda \hV$.
(This is the ``worst case'' scenario, as suggested by the finite-size scaling analysis in Sec.~\ref{subsec:finitesizescaling}.
Otherwise, the nonthermal persistence time is infinite.)
The eventual expectation value is then
\begin{align}
\lim_{t \to \infty} \la \Gamma(t)| \hat{m} |\Gamma(t) \ra = m_{\text{th.};\, \hH,\, E = \la \Gamma| \hH |\Gamma \ra} ~.
\end{align}
By perturbation theory in equilibrium quantum statistical mechanics, barring the unlikely situation where $\hH_0$ happens to have a first order transition at the temperature corresponding to average energy $E_0$, we expect that $m_{\text{th.};\, \hH,\, E = \la \Gamma| \hat{H} |\Gamma \ra}$ is close to $m_{\text{th.};\, \hH_0,\, E_0 = \la \Gamma| \hat{H}_0 |\Gamma \ra}$ in smallness in $\lambda$.
Hence, we expect that $\lim_{t \to \infty} \la \Gamma(t)| \hat{m} |\Gamma(t) \ra$ is a finite amount away from $\la \Gamma| \hat{m} |\Gamma \ra$.
Given the bound in Eq.~(\ref{eq:derivm}) on the time derivative, we have 
\begin{equation}
\big| \la \Gamma(t)| \hat{m} |\Gamma(t) \ra  - \la \Gamma| \hat{m} |\Gamma \ra \big| \leq \lambda (c_0 t + c_1 t^2/2) ~.
\label{eq:integrbound}
\end{equation}
Hence, at least until time $t^\star(\lambda) \sim \lambda^{-1/2}$ the observable will still be a finite amount away from the thermal value. 

Some remarks are in order here.
First and most importantly, while the bound Eq.~(\ref{eq:derivbound}) on the derivative of the observable is valid for any initial eigenstate, if $|\Gamma \ra$ were a ``thermal" state (i.e., satisfying ETH with respect to $\hH_0$), the above argument would not give us a divergent relaxation time for small $\lambda$ since in this case $\la \Gamma| \hat{m} |\Gamma \ra = m_{\text{th.};\, \hH_0,\, E_0 = \la \Gamma| \hat{H}_0 |\Gamma \ra}$ and is close to $\lim_{t \to \infty} \la \Gamma(t)| \hat{m} |\Gamma(t) \ra = m_{\text{th.};\, \hH,\, E = \la \Gamma| \hH |\Gamma \ra}$.
Thus, to obtain the divergent $t^\star(\lambda)$, it was crucial to use the nonthermal property of the scar states, namely that local observables have expectation values different from the thermal ones at the same energy density.

Second, we can relax the condition that the initial state $|\Gamma\ra$ is an eigenstate of $\hat{H}_0$ and only require that it produces nonthermalizing time evolution of the local observable under the unperturbed $\hat{H}_0$.
For example, this pertains to the persistent oscillations in some unperturbed models $\hat{H_0}$ starting with some special initial state $|\Gamma \ra$~\cite{moudgalyaExact2018, choiEmergent2019, schecterWeak2019}.

Indeed, using the following identity (which is a variant of Duhamel formula)
\begin{align}
\label{eq:Duhamel}
& e^{i \hat{H} t} \hat{m} e^{-i \hat{H} t} - e^{i \hat{H}_0 t} \hat{m} e^{-i \hat{H}_0 t} \\
& = i \int_0^t \!ds\,\, e^{i \hat{H} (t-s)} \left[ \lambda \hat{V}, e^{i \hat{H}_0 s} \hat{m} e^{-i \hat{H}_0 s} \right] e^{-i \hat{H} (t-s)} ~, \nonumber
\end{align}
we can bound
\begin{align}\label{eqn:oscillation_bound}
& \Big|\la \Gamma| e^{i \hat{H} t} \hat{m} e^{-i \hat{H} t} |\Gamma \ra - \la \Gamma| e^{i \hat{H}_0 t} \hat{m} e^{-i \hat{H}_0 t} |\Gamma \ra \Big| \nonumber \\
& \leq \int_0^t \!ds\,\, \Big\| \left[ \lambda \hat{V}, e^{i \hat{H}_0 s} \hat{m} e^{-i \hat{H}_0 s} \right] \Big\| \leq \lambda (c_0 t + c_1 t^2/2) ~. 
\end{align}
The integrand in the second line is bounded precisely as in Eq.~(\ref{eq:derivbound}) and leads to the final result.
Note that the above holds for any initial state $|\Gamma \ra$.
We are interested in situations when $|\Gamma \ra$ is still a special initial state such that $\la \Gamma| e^{i \hat{H}_0 t} \hat{m} e^{-i \hat{H}_0 t} |\Gamma \ra$ shows nonthermalizing time evolution, e.g., the persistent oscillations in the unperturbed model as happens in models with towers of exact scar states~\cite{moudgalyaExact2018, choiEmergent2019, schecterWeak2019}.
In this case, such nonthermal behavior will also be seen in the perturbed model at least until time of order $\lambda^{-1/2}$.
To further support our arguments, in Appendix~\ref{app:spin-1XY}, we show numerical results of the effect of perturbation on
perfect oscillation in the spin-1 XY model.

Third, the above arguments generalized to $d$ dimensions would replace the bound Eq.~(\ref{eq:derivbound}) on the derivative of the observable by $c_0 t^{d-1} + c_1 t^d$, which would yield thermalization time at least as long as $t^\star(\lambda) \sim \lambda^{-1/(d+1)}$.

\subsection{Possibility of stronger bounds on the thermalization time}\label{subsec:strongerbound}
We now ask if we can obtain stronger bounds on the thermalization time than the above $t^\star(\lambda) \sim \lambda^{-1/2}$ in $d = 1$.
The bound we obtained is indeed very general and used almost no information about $\hH_0$ and $\hV$.
The only assumptions we made were the locality of $\hH_0$ and $\hV$ and that $|\Gamma \ra$ violates ETH.

On the other hand, we suspect that in the problem of the PXP scar states, the thermalization time diverges with a stronger power law, perhaps even as $\lambda^{-1}$ in our specific model.
Such a suspicion can already be supported from Fig.~\ref{fig:tebd}, where we see that $|\la \Gamma(t)| \hat{m} |\Gamma(t) \ra - \la \Gamma| \hat{m} |\Gamma \ra|$ grows almost linearly in $t$ instead of the quadratic growth that appeared in the upper bound in Eq.~(\ref{eq:integrbound}). 
Equivalently, examination of the derivative $d\la \Gamma(t)| \hat{m} |\Gamma(t) \ra/dt$ of the measured observable in Fig.~\ref{fig:tebd} shows that the derivative remains bounded at least up to the time that the numerical calculation is reliable. 
Namely, there is no evidence of the linear in time growth that appears in the rigorous upper bound in Eq.~(\ref{eq:derivbound}), which suggests significant over-estimation in the bound.

More generally, the reason for this suspicion is as follows.
Consider the step in the argument in Sec.~\ref{subsec:rigorousbound} where for $|j| \leq v_{LR}\, t$, we used the bound $\| [\hat{v}_j, e^{i \hH t} \hat{m} e^{-i \hH t}] \| \leq 2 \|\hat{v}_j\| \|\hat{m}\|$.
While this is probably the best bound for the operator norm of the commutator, we are actually interested in the expectation value of the commutator evaluated in the initial state $|\Gamma \ra$.
This expectation value is essentially a retarded Green's function between the local operators $\hat{m}$ and $\hat{v}_j$, where the time dynamics is determined by $\hH$ and the initial ensemble is given by the pure state $|\Gamma \ra$.
The retarded Green's function in turn can be related to a dynamical time-ordered correlation function between the operators at space-time points $(0, t)$ and $(j, 0)$ (we take $t > 0$ throughout),
\begin{align}
-i \la \Gamma| [e^{i \hH t} \hat{m} e^{-i \hH t}, \hat{v}_j] |\Gamma \ra 
= 2 \text{Im} \left( \la \Gamma| e^{i \hH t} \hat{m} e^{-i \hH t} \hat{v}_j |\Gamma \ra \right) \notag ~,
\label{eq:GRmv}
\end{align}
where we have used hermiticity of $\hat{m}$ and $\hat{v}_j$.
It is likely to be very crude to use the operator norm to bound this expectation value.

Instead, we suspect that the above retarded Green's function decays both in space and in time, and the sum over $|j| \leq v_{LR}\, t$ can be bounded by a slower $t$-dependence than $t^1$ in Eq.~(\ref{eq:derivbound}), perhaps even by a $t$-independent number in many cases.
This suspicion is based on the physical intuition that correlation functions decay at large spatial or temporal separation (note that $\la \Gamma| e^{i \hH t} \hat{m} e^{-i \hH t} |\Gamma \ra$ and $\la \Gamma| \hat{v}_j |\Gamma \ra$ are both real numbers, so the above expression equals the imaginary part of the connected correlation function).
While for $|j| > v_{LR}\, t$ this expectation is formalized by the Lieb-Robinson bounds, here we need the regime $t > |j|/v_{LR}$, i.e., inside the ``light cone."

If $|\Gamma \ra$ were a thermal state and $\hH$ a thermalizing Hamiltonian, it would be natural to expect exponential decay of such correlations in time also inside the ``light cone." 
This would completely suppress the $t^1$ piece in Eq.~(\ref{eq:derivbound}).

As a less favorable example, suppose we know that the above correlation function is bounded by $\sim t^{-p}$ for all $|j| \leq v_{LR}\, t$.
Then in Eq.~(\ref{eq:derivbound}) we would be able to replace the right-hand-side bound by $\lambda (c_0 + c_1 t^{1-p})$ and in Eq.~(\ref{eq:integrbound}) the right-hand-side bound by $\lambda (c_0 t + c_1 t^{2-p}/(2-p))$.
If $p \geq 1$, then we could argue that the observable would still be a finite amount away from the thermal value at least until time $t^\star(\lambda) \sim \lambda^{-1}$, while if $p < 1$ we would only be able to argue that $t^\star(\lambda) \sim \lambda^{-1/(2-p)}$.
These conclusions would hold also if the retarded correlation function is bounded inside the light cone by $|G^R(x, t)| \leq A/(v_{LR}^2 t^2 - x^2)^{p/2}$, which is the form encountered at zero-temperature quantum critical points described by conformal field theories
(strictly speaking, we also need $p < 2$ for the spatial integral over $|x| < v_{LR}\, t$ to be convergent, where for concreteness we specialized to $d=1$).

From the above discussion, we see that if we allow arbitrary $\hH$ and $|\Gamma \ra$, we probably cannot improve the bound in Eq.~(\ref{eq:derivbound}) just on general grounds: the details of the system at hand are important.
However, for many systems it is likely that this bound significantly overestimates the actual rate of change of the observable, and the thermalization time will actually be significantly longer than suggested by the rigorous argument.
Our direct numerical study suggests that this is the case for the exact PXP scar states under the PXPZ perturbations.

\section{Conclusions}\label{sec:conclusions}
\subsection{Thermalization of exact scar states}
In this paper, we examine the fate of exact quantum many-body scar states under perturbations.
In particular, we consider the PXP model with two exact scar states at energies $E = \pm \sqrt{2}$, and two exact scars that are in the $E = 0$ degenerate manifold.
We consider the perturbation $\hat{V}$ in Eq.~(\ref{eq:VPXPZ}) that preserves the particle-hole property, the inversion symmetry, and hence the $E = 0$ manifold (but not the exact scars), to investigate the effects of the perturbation on the exact scar states.
We also compare the scar states to some thermal states under the perturbation.

In finite sizes, we find robust signatures of deformed scars in the perturbed Hamiltonian.
In particular, the smallness of the bipartite entanglement entropy and the presence of the VBS order seem to survive to some degree under the perturbation.
The hybridization of the scar states to other states also seems to be weaker compared to the hybridization of the thermal states, which is seen in the loss of fidelity relative to the unperturbed states.
Furthermore, we use Rayleigh-Schrodinger perturbation theory to construct perturbed scar states and find good agreement with the ED states.

Nevertheless, by examining the finite-size scaling of the off-diagonal matrix elements connecting the scar and thermal states, we conclude that the aforementioned robustness will be lost in the thermodynamic limit. 
In particular, we find that the off-diagonal matrix elements between the scar states and the nearby thermal states scale as ${\cal D}_L^{-1/2} \sim \phi^{-L/2}$, in agreement with the ETH picture of the thermal states.
Although these matrix elements have relatively smaller amplitudes compared to the off-diagonal matrix elements between thermal states, the difference appears to be quantitative and not qualitative.
Such a scaling of the matrix elements is not enough to beat the exponential decrease of the level-spacing going as ${\cal D}_L^{-1} \sim \phi^{-L}$, and accordingly we expect the eventual thermalization of the scar states.
We found similar results in our study of the perturbed spin-1 XY scar model in Appendix~\ref{app:spin-1XY}.
In addition, while not presented in this work explicitly, we also performed similar finite-size scaling analysis on the embedded Hamiltonian constructed in Ref.~\cite{okTopological}, and reached the same conclusions.
Accordingly, we expect that the eventual thermalization of the exact scar states under perturbations holds more generally.

On the other hand, despite the eventual thermalization, we show that the nonthermal properties of the exact scar states can survive for some parametrically long time even in the thermodynamic limit.
Specifically, we show numerically that the VBS order in the exact scar states in the PXP model can survive for a long time after a quench to the perturbed Hamiltonian.
We also present a general theory that rigorously lower-bounds the thermalization time as $t^{*} \sim O(\lambda ^{-1/(1+d)})$, where $d$ is the dimension of the system. 
In practice, depending on the details of the system, the thermalization time scale can even be longer.
By comparing the actual time evolution of the derivative of the observable in the numerical simulation with the bounds used in the rigorous argument, we propose that in the specific perturbed PXP model the thermalization time diverges at least as $\lambda^{-1}$.

Our work provides an important foundation for understanding stability of the exact scar states found and constructed in various special models, and possible relevance of such exact scar states for understanding apparent ``scarness'' of nearby models where exact scars are not known.
We are considering the ``worst case'' scenario, where an exact scar state in the spectrum is surrounded by chaotic states.
In this case, the random matrix theory description of the scaling of the off-diagonal matrix elements is perhaps unavoidable and leads to eventual thermalization.
However, the thermalization time can be large as long as the perturbation is small.
This means that in experiments, some nonthermal signatures in dynamics can indeed be understood from some special $\hat{H}_0$ and its exact scar states.
Furthermore, Ref.~\cite{shiraishiSystematic2017} proposed an ``embedded Hamiltonian" formalism as a way to engineer Hamiltonians which have nonthermal states inside the spectrum.
Our general lower bound on the thermalization time suggests that the nonthermal signatures of these states can survive for some long time even when perturbations break the exact embedded structure.
This also establishes the possibility of engineering embedded Hamiltonians to protect quantum information.

\subsection{Speculations on the origin of the numerical scars in the PXP model}
We conclude with some speculations about the PXP model itself, in light of our scenario of thermalization of scar states under generic perturbations.
Original ED studies~\cite{turnerQuantum2018, turnerWeak2018} found a band of prominent scar states in the model, while Ref.~\cite{linExact2019} found two exact scar states in periodic chains and four in open chains.
From our systematic study of the Schmidt numbers of all PXP eigenstates, we conjecture that only the latter can be expressed analytically for any system size, while the other numerically observed scars do not have exact analytic expressions.
One possible explanation is that the PXP model is proximate to some model that has a larger number of exact scars; and under the perturbation that takes this unknown model to the PXP model, only few of the original exact scars remain as exact eigenstates, while the rest do not.
If the perturbation is ``generic enough'' with respect to the states that do not remain exact, one could speculate based on the scar thermalization scenario that these scars will eventually thermalize, while their strong presence in ED is due to the relative weakness of the perturbation and limited system sizes.
It would certainly be interesting to look for such a tractable ``mother model" that could explain all scars that are very prominent in the PXP model.

However, some caution is in order about such speculation.
In the PXP model, Ref.~\cite{linExact2019} constructed single-mode approximation (SMA) and multi-mode approximation (MMA) states on top of the exact scar states that provide competitive approximations to the ED band of scars for the available system sizes $L$.
Although the SMA/MMA states are not close to the eigenstates as $L \to \infty$, our Apprndix~\ref{app:single_scar} shows that their thermalization time diverges with $L$.
This surprising breakdown of ETH for states that are orthogonal to the exact scar states suggests that some subtle non-thermalness survives in the spectrum excluding the exact scar states even in the thermodynamic limit.
For large sizes, this nonthermal property is likely spread over many eigenstates, since the SMA/MMA states can be expanded over eigenstates in a small energy window and this expansion should ``know'' about the divergent thermalization time of the trial states.
Finding precise diagnostics for such subtle non-thermalness at the level of eigenstates could be very challenging.

One may ask if the remaining subtle non-thermallness described above can be reconciled with our main story that scars thermalize under generic perturbations.
A possible explanation is that the perturbation envisioned here that takes one from the ``mother model'' to the PXP model is not the most general one since it has to retain the known exact scars whose presence is crucial to the above argument.

\begin{acknowledgments}
We thank Manuel Endres, Tarun Grover, Timothy Hsieh, Vedika Khemani, Michael Knap, Christopher Laumann, Zlatko Papi\'{c}, Brenden Roberts, Maksym Serbyn, Brian Timar, Christopher Turner, and Christopher White for valuable discussions.
This work was supported by National Science Foundation (NSF) through Grants No. DMR-1619696 (C.-J.~L. and O.~I.~M.) and DMR-1752759 (A.~C.). 
A.~C.\ further acknowledges support from the Sloan Foundation through the Sloan Research Fellowship.
C.-J.~L.\ acknowledges support from Perimeter Institute for Theoretical Physics.
Research at Perimeter Institute is supported in part by the Government of Canada through the Department of Innovation, Science and Economic Development Canada and by the Province of Ontario through the Ministry of Economic Development, Job Creation and Trade.
\end{acknowledgments}

\appendix
\section{Summary of the exact scar states and their properties in the PXP model}\label{app:summaryexactscar}

Reference~\cite{linExact2019} discovered four exact scar states in the PXP model $\hH_0$, Eq.~(\ref{eqn:PXP}), with open boundary conditions.
We denote the states as $|\Gamma_{\alpha\beta} \ra$, where $\alpha, \beta \in\{1, 2\}$.
The wavefunctions for $|\Gamma_{\alpha\beta} \ra$ are most economically expressed as matrix product states.

Defining the boundary vectors $v_1 = (1, 1)^T$ and $v_2 = (1, -1)^T$ and $2 \times 3$ and $3 \times 2$ matrices
\begin{eqnarray}
B^0 &=&
\begin{pmatrix}
1 & 0 & 0 \\
0 & 1 & 0
\end{pmatrix} ~, ~~~~~
B^1 = \sqrt{2}
\begin{pmatrix}
0 & 0 & 0 \\
1 & 0 & 1
\end{pmatrix} ~, \\
C^0 &=&
\begin{pmatrix}
0 & -1 \\
1 & 0 \\
0 & 0
\end{pmatrix} ~, ~~~~~
C^1 = \sqrt{2}
\begin{pmatrix}
1 & 0 \\
0 & 0 \\
-1 & 0
\end{pmatrix} ~,
\end{eqnarray}
we can express the scar states as
\begin{equation}
|\Gamma_{\alpha\beta} \ra = \frac{1}{\sqrt{N_{\alpha\beta}}} \sum_{\{\sigma\}} v_\alpha^T B^{\sigma_1} C^{\sigma_2} \dots B^{\sigma_{L-1}} C^{\sigma_L} v_\beta |\sigma_1 \dots \sigma_L \ra ~,
\label{eqn:GammaOBC}
\end{equation}
where $N_{\alpha\beta} = 2 \Big[3^{L/2} + (-1)^{L/2+\alpha+\beta} \Big]$ is the normalization factor.

The states $|\Gamma_{11} \ra$ and $|\Gamma_{22} \ra$ are not orthogonal but have overlap 
\begin{equation}
    \la \Gamma_{11}|\Gamma_{22} \ra = \frac{2}{3^{\frac{L}{2}} + (-1)^{\frac{L}{2}}} ~.
\end{equation}
The normalization factor of $|\Gamma_{I} \ra = \frac{1}{\sqrt{N}}( |\Gamma_{11} \ra - |\Gamma_{22} \ra)$ is therefore
\begin{equation}
    N = 2 - \frac{4}{3^{\frac{L}{2}} + (-1)^{\frac{L}{2}}} ~,
\end{equation}
as stated in the main text.

These scar states have the following symmetry properties.
For the inversion, we have 
\begin{align}
    I |\Gamma_{12} \ra &= -(-1)^{L/2} |\Gamma_{12} \ra ~, \\
    I |\Gamma_{21} \ra &= -(-1)^{L/2} |\Gamma_{21} \ra ~,\\
    I| \Gamma_{11} \ra &= (-1)^{L/2} |\Gamma_{22} \ra ~, \\
    I |\Gamma_{22} \ra &= (-1)^{L/2} |\Gamma_{11} \ra ~.
\end{align}
Hence $I |\Gamma_I \ra = -(-1)^{L/2} |\Gamma_I \ra$, i.e., $|\Gamma_I \ra$ has the same inversion quantum number as $|\Gamma_{21} \ra$.
For the particle-hole transformation, we have 
\begin{align}
    \mathcal{C} |\Gamma_{12} \ra &= (-1)^{L/2} |\Gamma_{21} \ra ~, \\
    \mathcal{C} |\Gamma_{11} \ra &= (-1)^{L/2} |\Gamma_{22} \ra ~.
\end{align}

\section{Proof of the perturbation energy $E_{\Gamma_I}^{(n)}=0$ to all orders}\label{app:EGammaIallorder}
In this appendix, we prove that the perturbation energy for $|\Gamma_I \ra \equiv |\Gamma_I^{(0)} \ra$ gives $E_{\Gamma_I}^{(n)}=0$ to all orders.
The proof is obtained via mathematical induction, by showing that the perturbative wavefunction $|\Gamma_I^{(n)} \ra$ is an eigenstate of the particle-hole transformation with eigenvalue $\mathcal{C} = -(-1)^{L/2}$ to any order.
(We caution, however, that this does not imply that the perturbation theory converges.)

First, we briefly review the formal development of the Rayleigh-Schrodinger perturbation expansion. 
We closely follow the notations and settings in Ref.~\cite{sakurai_napolitano_2017}.
In the perturbation theory, we are trying to solve the eigenvalue equation
\begin{equation}\label{eqn:pert_eigequation}
    (E_{\Gamma_I}^{(0)} - H_0)|\Gamma_I(\lambda) \ra = [\lambda V - \Delta_{\Gamma_I}(\lambda)] |\Gamma_I(\lambda) \ra ~,
\end{equation}
where $\Delta_{\Gamma_I}(\lambda) = E_{\Gamma_I}(\lambda) - E_{\Gamma_I}^{(0)}$ is the energy shift. 
Note that we can immediately see $\la \Gamma_I^{(0)}| (\lambda V -\Delta_{\Gamma_I}) |\Gamma_I(\lambda) \ra = 0$, or 
\begin{equation}\label{eqn:Deltafull}
    \Delta_{\Gamma_I}(\lambda) = \la \Gamma_I^{(0)}| \lambda V |\Gamma_I(\lambda) \ra ~.
\end{equation}
Assuming we have chosen the basis such that $\la E_n^{(0)}\!=\!0|V| E_m^{(0)}\!=\!0\ra =0$ for $m \neq n$ in the $E=0$ degenerate manifold, by defining $\phi_{\Gamma_I}=I-\sum_{k:E_k^{(0)}=0}|k^{(0)}\ra\la k^{(0)}|$, we have a formal solution for Eq.~(\ref{eqn:pert_eigequation}):
\begin{equation}\label{eqn:formalsolution}
    |\Gamma_I(\lambda)\ra = |\Gamma^{(0)}\ra + \frac{\phi_{\Gamma_I}}{E_{\Gamma_I}^{(0)}-H_0}[\lambda V - \Delta_{\Gamma_I} (\lambda)]|\Gamma_I(\lambda)\ra~.
\end{equation}

We further assume the (formal) series expansion of $|\Gamma_I(\lambda)\ra = \sum_{n=0}^{\infty}\lambda^n |\Gamma_I^{(n)}\ra$ and $\Delta_{\Gamma_I}(\lambda) = \sum_{m=1}^{\infty}\lambda^{m}\Delta_{\Gamma_I}^{(m)}$.
From Eq.~(\ref{eqn:Deltafull}) and plugging in the series expansion of $\Delta_{\Gamma_I}(\lambda)$, equating the same order of $\lambda$, we have, at $N$-th order
\begin{equation}\label{eqn:Delta_Nth}
    \Delta_{\Gamma_I}^{(N)}=\la \Gamma_{I}^{(0)}|V|\Gamma_I^{(N-1)}\ra.
\end{equation}
Moreover, from Eq.~(\ref{eqn:formalsolution}) and plugging in the series expansion of $|\Gamma_I(\lambda)\ra$, equating the same order of $\lambda$, we have, at $N$-th order
\begin{align}\label{eqn:Gamma_Nth}
    |\Gamma_I^{(N)}\ra&=\frac{\phi_{\Gamma_I}}{E^{(0)}_{\Gamma_I}-H_0}V|\Gamma_I^{(N-1)}\ra \notag \\
    &- \frac{\phi_{\Gamma_I}}{E^{(0)}_{\Gamma_I}-H_0} \sum_{n=0}^{N-1}\Delta_{\Gamma_I}^{(N-n)}|\Gamma_I^{(n)}\ra~.
\end{align}
Eqs.~(\ref{eqn:Delta_Nth}) and (\ref{eqn:Gamma_Nth}) are recursive: the former requires only $|\Gamma_I^{(N-1)} \ra$ while the latter requires $|\Gamma_I^{(m)} \ra$ with $m = 0, \dots, N-1$ and $\Delta_{\Gamma_I}^{(n)}$ with $n = 1, \dots, N$ (where $n=N$ was just calculated).

Now we have the essential ingredients for mathematical induction.
Recall that we are interested in the type of the perturbation $V$ having the property $\mathcal{C}\hV=-\hV\mathcal{C}$, and also $\mathcal{C}|\Gamma_I^{(0)}\ra=-(-1)^{L/2}|\Gamma_I^{(0)}\ra$ hence $E_{\Gamma_I}^{(0)}=0$.
Assume $|\Gamma_I^{(m)} \ra$, where $m=0 \dots N-1$ all have the particle-hole quantum number $\mathcal{C}=-(-1)^{L/2}$.
We can immediately see from Eq.~(\ref{eqn:Delta_Nth}) that $\Delta_{\Gamma_I}^{(n)}=0$ for $n=1 \dots N $.
Accordingly, we have the perturbed wavefunction at $N$-th order as
\begin{widetext}
\begin{align}
    |\Gamma_I^{(N)}\ra &= \frac{\phi_{\Gamma_I}}{E^{(0)}_{\Gamma_I}-H_0}V|\Gamma_I^{(N-1)}\ra 
    = \sum_{k:E_k^{(0)} > 0} \left ( -\frac{|k^{(0)}\ra\la k^{(0)}|V|\Gamma_I^{(N-1)} \ra}{E_k^{(0)}} + \frac{\mathcal{C}|k^{(0)}\ra\la k^{(0)}|\mathcal{C}V|\Gamma_I^{(N-1)} \ra}{E_k^{(0)}} \right) \notag \\
    & = \sum_{k:E_k^{(0)} > 0} \left ( -\frac{|k^{(0)}\ra\la k^{(0)}|V|\Gamma_I^{(N-1)} \ra}{E_k^{(0)}} +(-1)^{L/2} \frac{\mathcal{C}|k^{(0)}\ra\la k^{(0)}|V|\Gamma_I^{(N-1)} \ra}{E_k^{(0)}} \right)~.
\end{align}
\end{widetext}

Therefore, $\mathcal{C}|\Gamma_I^{(N)} \ra = -(-1)^{L/2}|\Gamma_I^{(N)} \ra$. 
By mathematical induction, we have $\mathcal{C}|\Gamma_I^{(n)} \ra = -(-1)^{L/2}|\Gamma_I^{(n)} \ra$ to any order.
This indeed also implies $\Delta_{\Gamma_I}^{(n)}=0$ for any order $n$.

\section{Diagonally improved perturbation theory}\label{app:diag_pert_theory}
To further elaborate on the use of the perturbation theory in finite sizes in Sec.~\ref{subsec:perturb} when there are avoided level-crossings, particularly for the scar state $|\Gamma_{21} \ra$, we modify the perturbation theory in the following way.
We include in the ``unperturbed'' Hamiltonian the diagonal part of $\hV$, and treat the off-diagonal part of $\hV$ as perturbation. 
More specifically, in the eigenbasis of $\hH_0$, namely $|n^{(0)} \ra$, the matrix element of the diagonal part of $\hV$ is $[\hV_{\text{diag.}}]_{nm} = \la n^{(0)}| \hV |n^{(0)} \ra \delta_{nm} $ and the off-diagonal part is $\hV_{\text{off-diag.}} = \hV - \hV_{\text{diag.}}$.
The unperturbed Hamiltonian is now $\hH_0' = \hH_0 + \lambda \hV_{\text{diag.}}$ and the perturbation is $\lambda \hV' = \hV_{\text{off-diag.}}$.
Clearly, the unperturbed eigenstates are still $|n^{(0)} \ra$.

Based on this regrouping, we can use the standard perturbation theory Eqs.~(\ref{eqn:perturbation_energy}) and (\ref{eqn:pert_fidelity}) with the replacements 
\begin{align}
    E_{n}^{(0)} & \rightarrow E_{n}^{(0)} + \lambda \, \la n^{(0)}| \hV_{\text{diag.}} |n^{(0)} \ra \notag ~, \\ 
    \hV & \rightarrow \hV_{\text{off-diag.}} ~. \notag
\end{align}

\begin{figure}
    \includegraphics[width=\columnwidth]{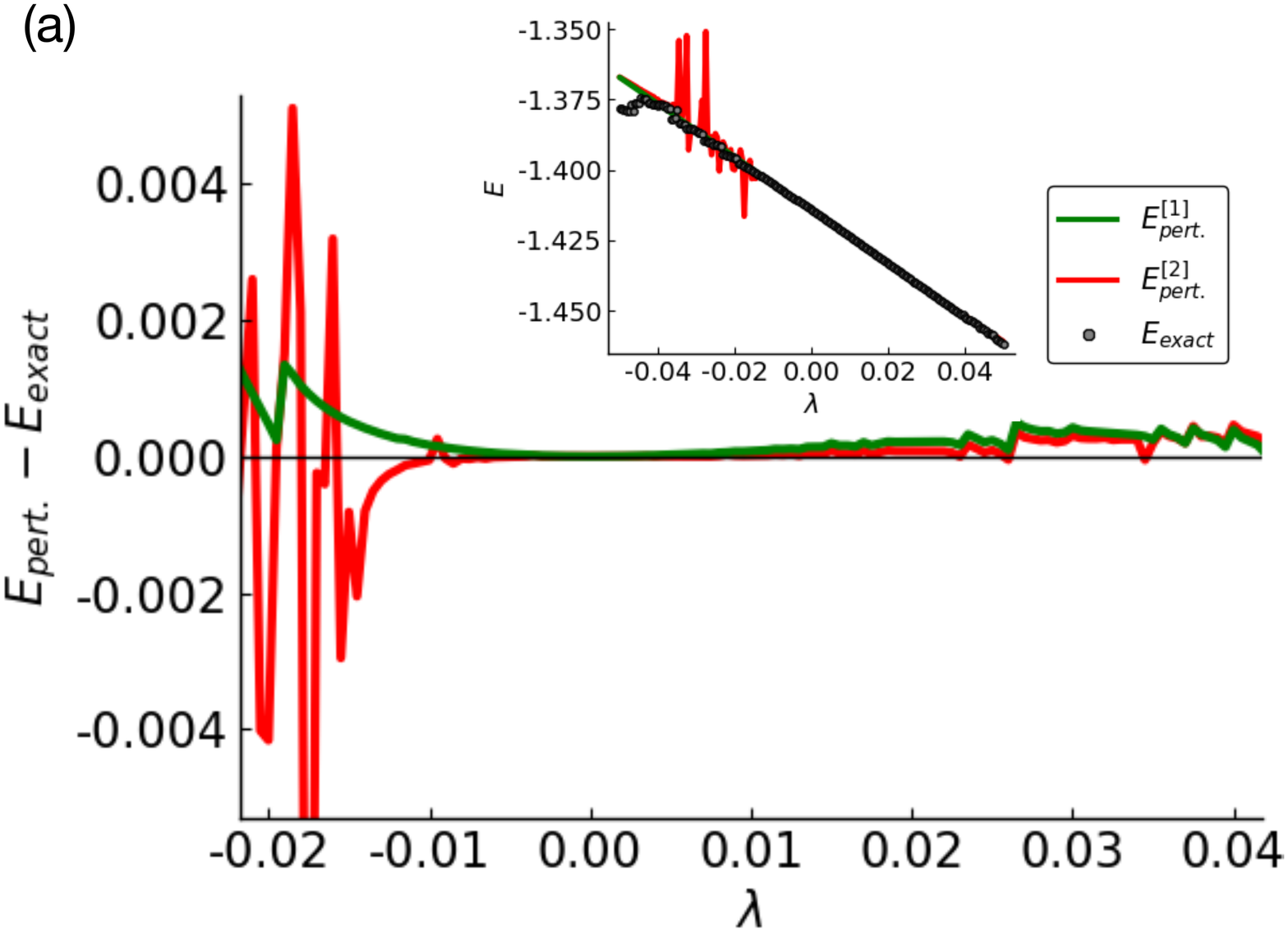}
    \includegraphics[width=\columnwidth]{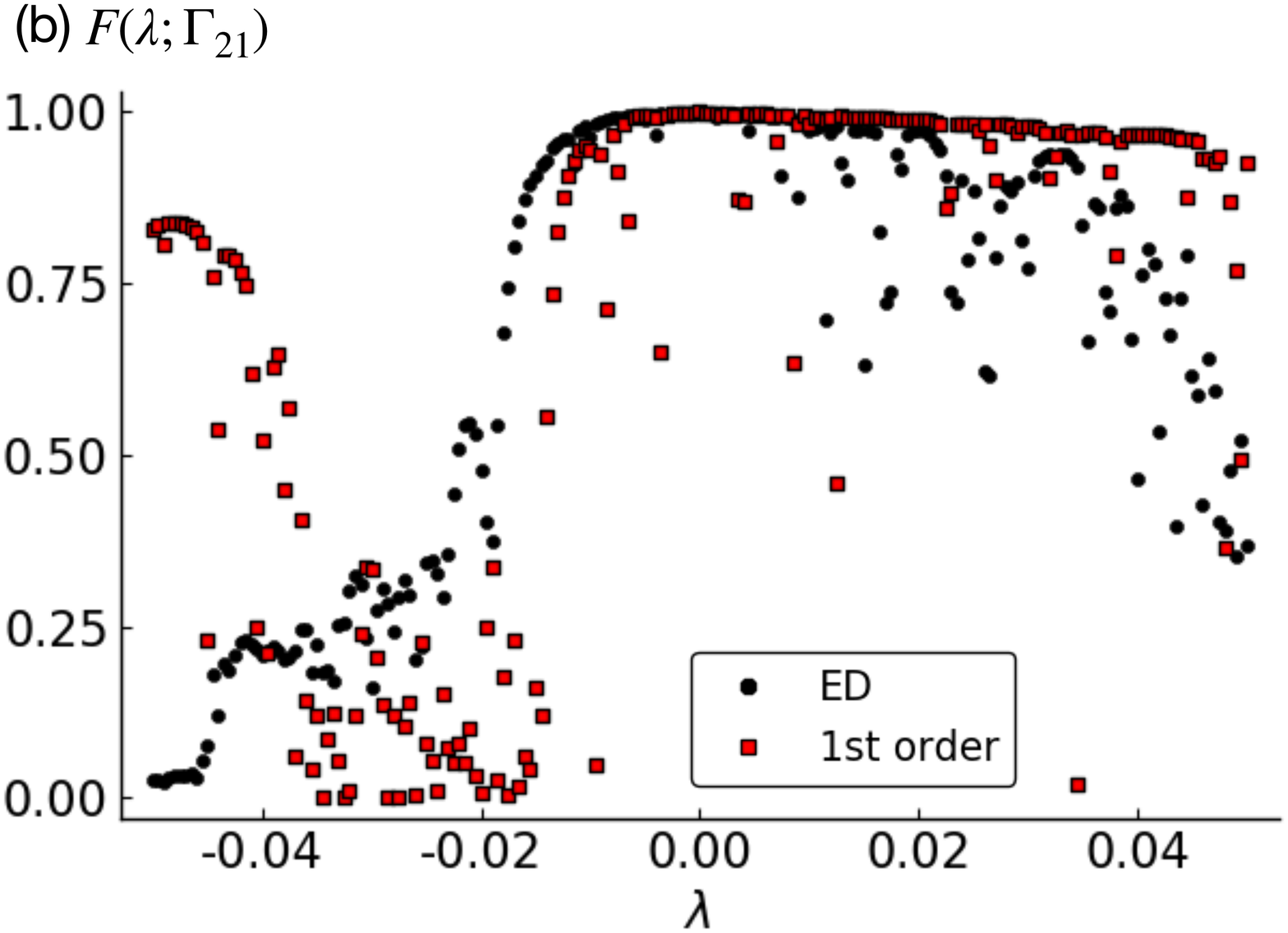}
    \caption{(a) Comparison of the diagonally improved perturbation energy and the ED energy of the perturbed $|\Gamma_{21} \ra$.
    (b) Comparison of the fidelity obtained in the diagonally improved perturbation theory and the ED result. 
    The perturbation theory now also accounts for the phenomenon of accidental hybridization.} 
    \label{fig:dagonal_perturbation}
\end{figure}

In Figs.~\ref{fig:dagonal_perturbation}(a) and (b), we show the comparison between the diagonally improved perturbation theory and the ED results for the state $|\Gamma_{21} \ra$, similar to Figs.~\ref{fig:EnergyPert} and \ref{fig:FidelityPert}.
Note that on the $\lambda < 0$ side, the perturbation theory also shows a drop of the overlap around $\lambda \approx -0.02$, though the recovery at $\lambda \approx -0.04$ is spurious.
On the $\lambda > 0 $ side, the diagonally improved perturbation theory also shows the accidental hybridization phenomenon. 
It happens when $E^{(0)}_n + \lambda \la n^{(0)}| \hV |n^{(0)} \ra - E^{(0)}_{\Gamma_{21}} - \lambda \la \Gamma_{21}| \hV |\Gamma_{21} \ra$ happens to be small compared to the matrix element $\la n^{(0)}| \hV |\Gamma_{21} \ra$.

We can now appreciate the improvement provided by this perturbation theory but also its eventual failure.
For an isolated level crossing driven by increasing $\lambda$ and the difference between $\lambda \la n^{(0)}| \hV |n^{(0)} \ra$ and $\lambda \la \Gamma_{21}| \hV |\Gamma_{21} \ra$, this perturbation theory is accurate both well before the crossing and well after the crossing, which is why we see improvement in the 2nd-order estimate of the energy in Fig.~\ref{fig:dagonal_perturbation}(a) compared to Fig.~\ref{fig:EnergyPert}, particularly on the $\lambda > 0$ side.
However, once the separation between successive energy level crossings (set by the density of states) becomes smaller than the duration of the avoided level crossings (set by the off-diagonal matrix elements), the improved perturbation theory also fails.
Upon increasing the system size, the density of states and hence the frequency of level crossings increases very fast and cannot be compensated by the decrease of the typical off-diagonal matrix elements, as discussed in Sec.~\ref{subsec:finitesizescaling}.
Hence the failure of the perturbation theory and the eventual thermalization of the scar states appear inevitable in the thermodynamic limit.
In the perturbed PXP model here, for the perturbation strength $\lambda \simeq 0.05$, our largest ED system sizes are just reaching the regime where this starts to happen.

\section{Some properties of the distribution of the matrix elements $\la n^{(0)}| \hV |\Gamma \ra$ as dictated by the locality of $\hH_0$ and $\hV$}
\label{app:distr_matr_elem}
Here we collect some observations about the distribution of the matrix elements $\la n^{(0)}| \hV |\Gamma \ra$.
As a simple application, we then demonstrate how the naive 2nd-order time-independent and time-dependent perturbation theory approaches fail in the thermodynamic limit for any fixed $\lambda \neq 0$.

First, we note that
\begin{align}
\sum_{n \neq \Gamma} |\la n^{(0)}| \hV |\Gamma \ra|^2 = \la \Gamma| \hV^2 |\Gamma \ra - \la \Gamma| \hV |\Gamma \ra^2 \equiv \text{var}(\hV; \Gamma) ~,
\label{eq:varVdef}
\end{align}
where all states are assumed normalized (here and below, we use the same notation as in Sec.~\ref{subsec:perturb}).
This allows us to think about the squared amplitudes of the matrix elements between the $|\Gamma \ra$ and other states $|n^{(0)} \ra$ as some ``weights,'' where the total weight is given by the variance of the perturbation $\hV$ in the state $|\Gamma \ra$.
Since $\hV$ is a local Hamiltonian, $\hV = \sum_j \hat{v}_j$, the variance of $\hV$ grows linearly with the system size $L$:
\begin{align}
\text{var}(\hV; \Gamma) & = \sum_{j,j'} \left[ \la \Gamma| \hat{v}_j \hat{v}_{j'} |\Gamma \ra - \la \Gamma| \hat{v}_j |\Gamma \ra \la \Gamma| \hat{v}_{j'} |\Gamma \ra \right] \nonumber \\
& = \sum_j \sum_{j'} G_{vv; \Gamma}(j-j') \approx \alpha L ~.
\label{eq:varValphaL}
\end{align}
Here we have assumed translational invariance and that connected correlation functions of local observables are short-range, which is true for our exact scar states with the finite bond dimension.
We expect this to be true also for short-range-correlated thermal states (e.g., away from any finite-temperature critical point or critical phase), in particular for the thermal states used in the main text.

\begin{figure}
    \includegraphics[width=\columnwidth]{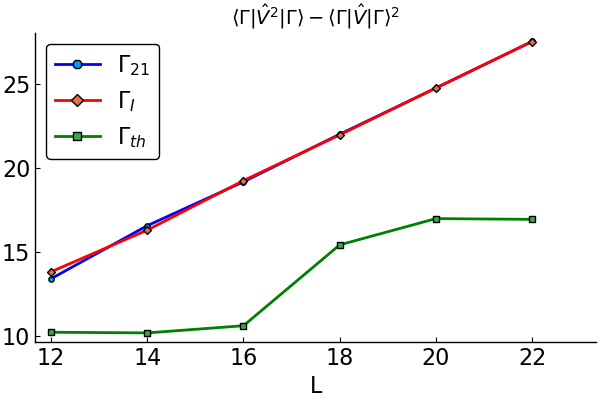}
    \caption{System size scaling of the variance of $\hV$ in eigenstates $|\Gamma_{21} \ra$, $|\Gamma_I \ra$, and $|\Gamma_{\text{th}} \ra$.
    Since the thermal state $|\Gamma_{\text{th}} \ra$ is inherently random, its system size dependence is ``noisy." }
    \label{fig:VFlucScaling}
\end{figure}

Figure~\ref{fig:VFlucScaling} shows measured variance of $\hV$ in the scar and thermal states used in the main text as a function of $L$ and confirms the expected linear in size scaling.
The plot for the thermal state has noisy $L$ dependence, which is expected since the thermal states are inherently ``random," so even our ``deterministic" procedure of picking the thermal state whose index is given by the index of the exact scar state plus three, when going from one size to the next has effective randomness in it.
On the other hand, the plots for the scar states use the same MPS but for different systems sizes, and there is no randomness in this.

We next observe that the mixing weights $|\la n^{(0)}| \hV |\Gamma \ra|^2$ are significant only for states with $|E_n^{(0)} - E_\Gamma^{(0)}| \lesssim O(1)$.
The reason is again locality of $\hV$:
Each $\hat{v}_j$ when acting on $|\Gamma \ra$ can ``add" or ``remove" only $O(1)$ energy as measured by $\hat{H}_0$, so $\hat{v}_j |\Gamma \ra$ expanded over the eigenstates of $\hat{H}_0$ is concentrated on $|n^{(0)} \ra$ with $|E_n^{(0)} - E_\Gamma^{(0)}| \lesssim O(1)$.
This observation can be formalized as follows.
Consider
\begin{align}
&\sum_{n \neq \Gamma} \left( E_n^{(0)} - E_\Gamma^{(0)} \right)^2 |\la n^{(0)}| \hV |\Gamma \ra|^2 
= \sum_{n \neq \Gamma} |\la n^{(0)}| [\hH_0, \hV] |\Gamma \ra|^2 \nonumber \\
& = \la \Gamma| \hat{W}^2 |\Gamma \ra - \la \Gamma| \hat{W} |\Gamma \ra^2 
\equiv \text{var}(\hat{W}; \Gamma) ~.
\label{eq:varWdef}
\end{align}
Here
\begin{align}
\hat{W} \equiv i [\hH_0, \hV] = \sum_j i [\hH_0, \hat{v}_j] = \sum_j \hat{w}_j
\end{align}
is an operator which is a sum of local terms $\hat{w}_j$ (the factor of $i = \sqrt{-1}$ makes $\hat{W}$ hermitian to simplify expressions).
Hence, similarly to Eq.~(\ref{eq:varValphaL}), the variance of $\hat{W}$ in the state $\Gamma$ grows linearly with the system size,
\begin{align}
\text{var}(\hat{W}; \Gamma) = \sum_j \sum_{j'} G_{ww; \Gamma}(j - j') \approx \beta L ~.
\label{eq:varWbetaL}
\end{align}
We then conclude that for large $L$
\begin{align}
\frac{\sum_{n \neq \Gamma} \left( E_n^{(0)} - E_\Gamma^{(0)} \right)^2 |\la n^{(0)}| \hV |\Gamma \ra|^2}{\sum_{n \neq \Gamma} |\la n^{(0)}| \hV |\Gamma \ra|^2} \approx \frac{\beta}{\alpha} ~.
\label{eq:varEvarW}
\end{align}
The left hand side can be interpreted as an average of $\left( E_n^{(0)} - E_\Gamma^{(0)} \right)^2$ over the distribution of $n$'s with weights proportional to $|\la n^{(0)}| \hat{V} |\Gamma \ra|^2$, and we see that this average is an $L$-independent number, as claimed.

Figure~\ref{fig:offdiag} in the main text indeed shows that the matrix elements $|\la n^{(0)}| \hat{V} |\Gamma \ra|$ become very small once $|E_n^{(0)} - E_\Gamma^{(0)}|$ exceeds some characteristic energy scale.
This is true for both the scar and thermal states $|\Gamma \ra$ studied.

\subsection{Simple-minded application to the time-independent 2nd-order perturbation theory and the large $L$ limit}
As an application, consider the formal 2nd-order perturbation theory expression for the ratio of the probability of being in any state other than $\Gamma$ to the probability of remaining in the state $\Gamma$ [cf.\ Eq.~(\ref{eqn:pert_fidelity}) in the main text]:
\begin{align}
R \equiv \sum_{n \neq \Gamma} \frac{|\la n^{(0)}| \lambda \hV |\Gamma \ra|^2}{\left( E_n^{(0)} - E_\Gamma^{(0)} \right)^2} ~.
\label{eqn:R}
\end{align}
We can bound this as follows:
\begin{align}
R &\geq \frac{\lambda^2}{\Delta^2} \sum_{n \neq \Gamma,~ |E_n^{(0)} - E_\Gamma^{(0)}| \leq \Delta} |\la n^{(0)}| \hV |\Gamma \ra|^2 \nonumber \\
&= \frac{\lambda^2}{\Delta^2} \left[ \text{var}(\hV; \Gamma) - \sum_{n,~ |E_n^{(0)} - E_\Gamma^{(0)}| > \Delta} |\la n^{(0)}| \hV |\Gamma \ra|^2 \right] ~. \nonumber
\end{align}
Here we used Eq.~(\ref{eq:varVdef}) and introduced some fixed energy scale $\Delta$, which will be chosen below.
We can now use Eq.~(\ref{eq:varWdef}) to bound the second term in the square brackets as follows:
\begin{align}
\text{var}(\hat{W}; \Gamma) & \geq \sum_{n,~ |E_n^{(0)} - E_\Gamma^{(0)}| > \Delta} \left( E_n^{(0)} - E_\Gamma^{(0)} \right)^2 |\la n^{(0)}| \hV |\Gamma \ra|^2 \nonumber \\
& \geq \Delta^2 \sum_{n,~ |E_n^{(0)} - E_\Gamma^{(0)}| > \Delta}
|\la n^{(0)}| \hV |\Gamma \ra|^2 ~.
\label{eq:varWbound}
\end{align}
We finally obtain
\begin{align}
R &\geq \frac{\lambda^2}{\Delta^2} \left[ \text{var}(\hV; \Gamma) - \frac{1}{\Delta^2} \text{var}(\hat{W}; \Gamma) \right] \nonumber \\
& \approx \frac{\lambda^2}{\Delta^2} \left[\alpha - \frac{\beta}{\Delta^2} \right] L = \frac{\lambda^2 \alpha^2}{4\beta} L ~,\quad \text{for~} \frac{1}{\Delta^2} = \frac{\alpha}{2\beta} ~,
\label{eq:staticbound}
\end{align}
where in the very last equation we picked $\Delta$ to make this lower bound as large as possible.

Thus, we see that the probability for the perturbed eigenstate to remain in the initial state $|\Gamma \ra$ decreases to zero with $L$, at least in this formal treatment.
Of course, it is known that the fidelity of a many-body state under a generic perturbation goes to zero in the thermodynamic limit, and the above is not intended as any serious proof but only as a simple illustration of thinking about the distribution of the matrix elements.
Note that the bound does not use any information about this distribution except the variance; in particular, it applies also, e.g., for a gapped ground state that is separated from the rest of the states by a gap---the fidelity under perturbation still goes to zero.
In the case of states at finite energy density that are surrounded by many thermal states, for very large $L$ the above lower bound is actually a gross underestimate of how poorly the formal static second-order perturbation theory performs:
With the level spacing decreasing as $\sim {\cal D}_L^{-1}$ (where ${\cal D}_L$ is the dimension of the total Hilbert space which grows exponentially with $L$), and the matrix elements decreasing as $\sim \mathcal{D}_L^{-1/2}$, the individual terms $\frac{|\la n^{(0)}| \lambda \hV |\Gamma \ra|^2}{\left( E_n^{(0)} - E_\Gamma^{(0)} \right)^2}$ associated with levels that are next to the $|\Gamma \ra$ level increase as $\sim \mathcal{D}_L$, which is much faster than the linear in $L$ lower bound in Eq.~(\ref{eq:staticbound}).

\subsection{Application to the time-dependent 2nd-order perturbation theory for fidelity after a quench}
As another application, consider the following quantity:
\begin{align}
P(t) \equiv \sum_{n \neq \Gamma} \frac{4 |\la n^{(0)}| \lambda \hV |\Gamma \ra|^2}{\left( E_n^{(0)} - E_\Gamma^{(0)} \right)^2} \sin^2 \frac{\left( E_n^{(0)} - E_\Gamma^{(0)} \right) t}{2} ~,
\label{eq:Pexact_in_perttheory}
\end{align}
which arises when we apply the time-dependent 2nd-order perturbation theory to the quench setting described in Sec.~\ref{sec:prethermalization}.
Specifically, the system starts in the state $|\Gamma \ra$ at time $t = 0$ and evolves under the perturbed Hamiltonian $\hH = \hH_0 + \lambda \hV$.
The above $P(t)$ is the perturbative result for the probability of being in any state other than $\Gamma$ at time $t$; hence, the probability of remaining in the state $\Gamma$, or fidelity at time $t$ after the quench, is $1 - P(t)$.
Clearly, $P(t)$ also lower-bounds the quantity $R$ in Eq.~(\ref{eqn:R}) that arises in the time-independent 2nd-order perturbation theory.
However, the dynamical quench setting is more interesting in that the stated results are potentially more accurate and useful on reasonable time or length scales because of the effective control over the denominators provided by the $\sin^2 \frac{(E_n^{(0)} - E_\Gamma^{(0)}) t}{2}$ factors.

For very small $t$ such that for all significantly-participating $n$ the quantity $(E_n^{(0)} - E_\Gamma^{(0)}) t$ is small, we have
\begin{align}
P(t) \approx t^2 \lambda^2 \, \text{var}(\hV; \Gamma) \approx t^2 \lambda^2 \alpha L ~.
\label{eq:Papprox}
\end{align}
We know from the preceding discussion that the weights $|\la n^{(0)}| \hat{V} |\Gamma \ra|^2$ are significant only if $|E_n^{(0)} - E_\Gamma^{(0)}| \lesssim O(1)$, so we actually expect the above formula to be an accurate description of the formal perturbation theory result up to $t \sim O(1)$.
In particular, this formula is valid for $t \sim 1/\sqrt{\lambda^2 \alpha L}$, beyond which the perturbation theory would give the probability $P(t)$ exceeding unity.
Beyond this time, we clearly cannot apply such a formulation of the perturbation theory, but until a somewhat smaller time of the same order such an application actually appears sensible.

More precisely, let us pick a number $y \in (0, \pi)$ and set $f_y = \sin^2(y)/y^2$.
We have the following bounds:
\begin{align}
& P(t) \geq \lambda^2 t^2 f_y \sum_{n \neq \Gamma,~ |E_n^{(0)} - E_\Gamma^{(0)}| \leq 2 y / t} |\la n^{(0)}| \hV |\Gamma \ra|^2 \nonumber \\
& = \lambda^2 t^2 f_y 
\bigg[\sum_{n \neq \Gamma} |\la n^{(0)}| \hV |\Gamma \ra|^2 - \!\!\!\!\!\!\! \sum_{n,~ |E_n^{(0)} - E_\Gamma^{(0)}| > 2 y / t} \!\!\!\!\!\!\!\! |\la n^{(0)}| \hV |\Gamma \ra|^2 \bigg] \nonumber \\
& \geq \lambda^2 t^2 f_y \left[\text{var}(\hV; \Gamma) - \frac{t^2}{4 y^2} \text{var}(\hat{W}; \Gamma) \right] \nonumber \\
& \approx \lambda^2 t^2 f_y L \left[\alpha - \frac{t^2}{4 y^2} \beta \right] ~,
\label{eq:Pbound}
\end{align}
where in the third line we used Eqs.~(\ref{eq:varVdef}) and (\ref{eq:varWbound}), while in the last line we used Eqs.~(\ref{eq:varValphaL}) and (\ref{eq:varWbetaL}).
For simplicity, let us keep $O(1)$ number $y$ and hence $f_y$ fixed, i.e., we do not try to further optimize this degree of freedom.
We see that, e.g., for $t < y \sqrt{\alpha/\beta}$ and $t > 1/\sqrt{3 \lambda^2 f_y L \alpha/4}$, which can be satisfied simultaneously for large enough $L$, we have $P(t) > 1$.
Hence, the 2nd-order perturbation theory already fails beyond a time that scales as $L^{-1/2}$, as claimed earlier.

To summarize, the above analysis suggests that the time scale for the fidelity loss goes to zero for large $L$ but only as a power law $L^{-1/2}$.
We remark that the above arguments are true for both scar and thermal states, and the difference in such short-time fidelity loss is only quantitative.
Nevertheless, the above analysis helps, e.g., when we want to understand ED results for the fidelity loss in the quench setting of Sec.~\ref{sec:prethermalization}.
We did not present such fidelity results as they do not allow defining slow thermalization in the thermodynamic limit.
Instead, as presented in the main text, measuring local observables shows that the nonthermal signatures of the scar in the initial state persist to nonzero time even when $L \to \infty$, and this thermodynamic-limit time diverges when the perturbation strength goes to zero.

\section{Effect of perturbations on scar states in the spin-1 XY model}
\label{app:spin-1XY}
\begin{figure}
    \includegraphics[width=\columnwidth]{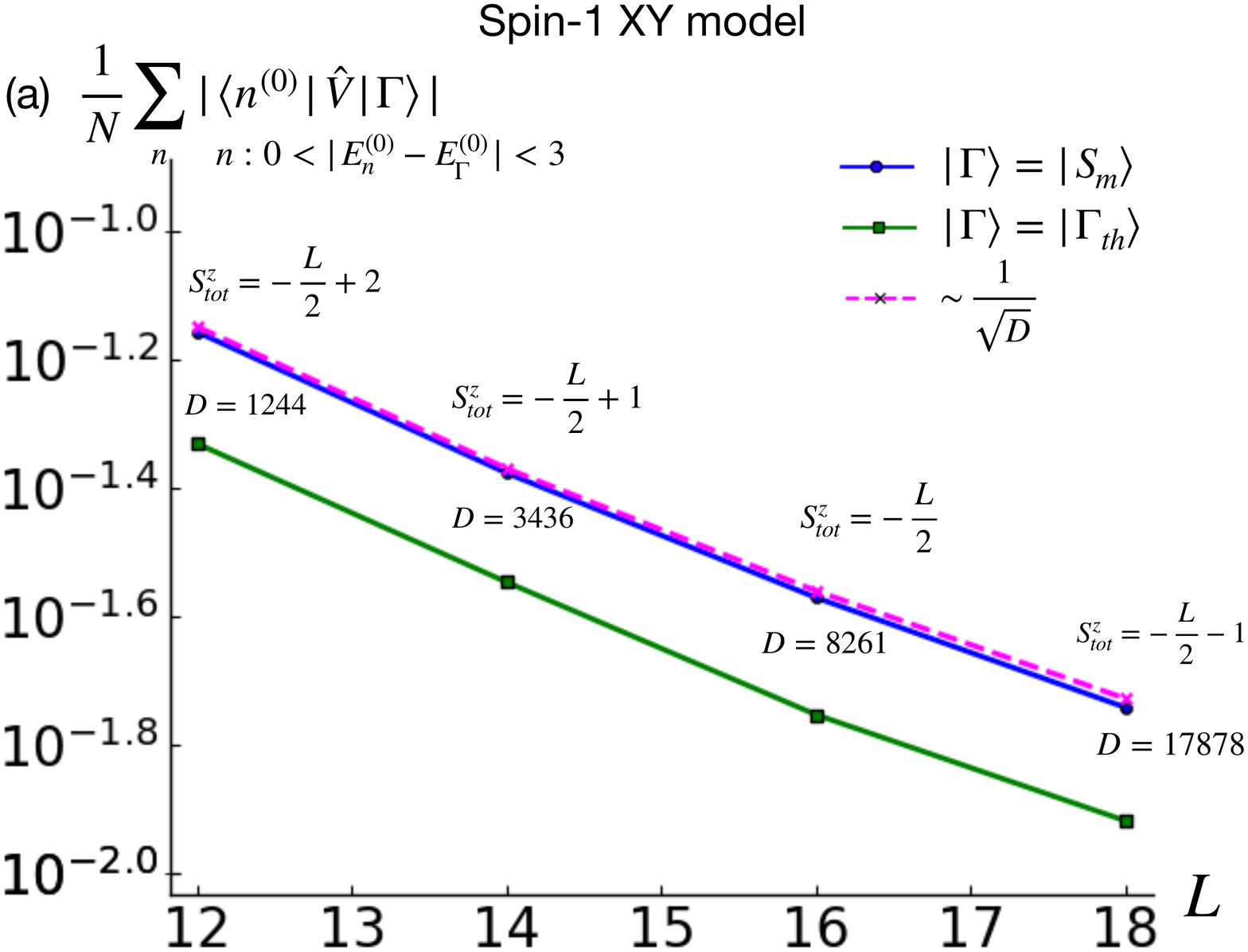}
    \includegraphics[width=\columnwidth]{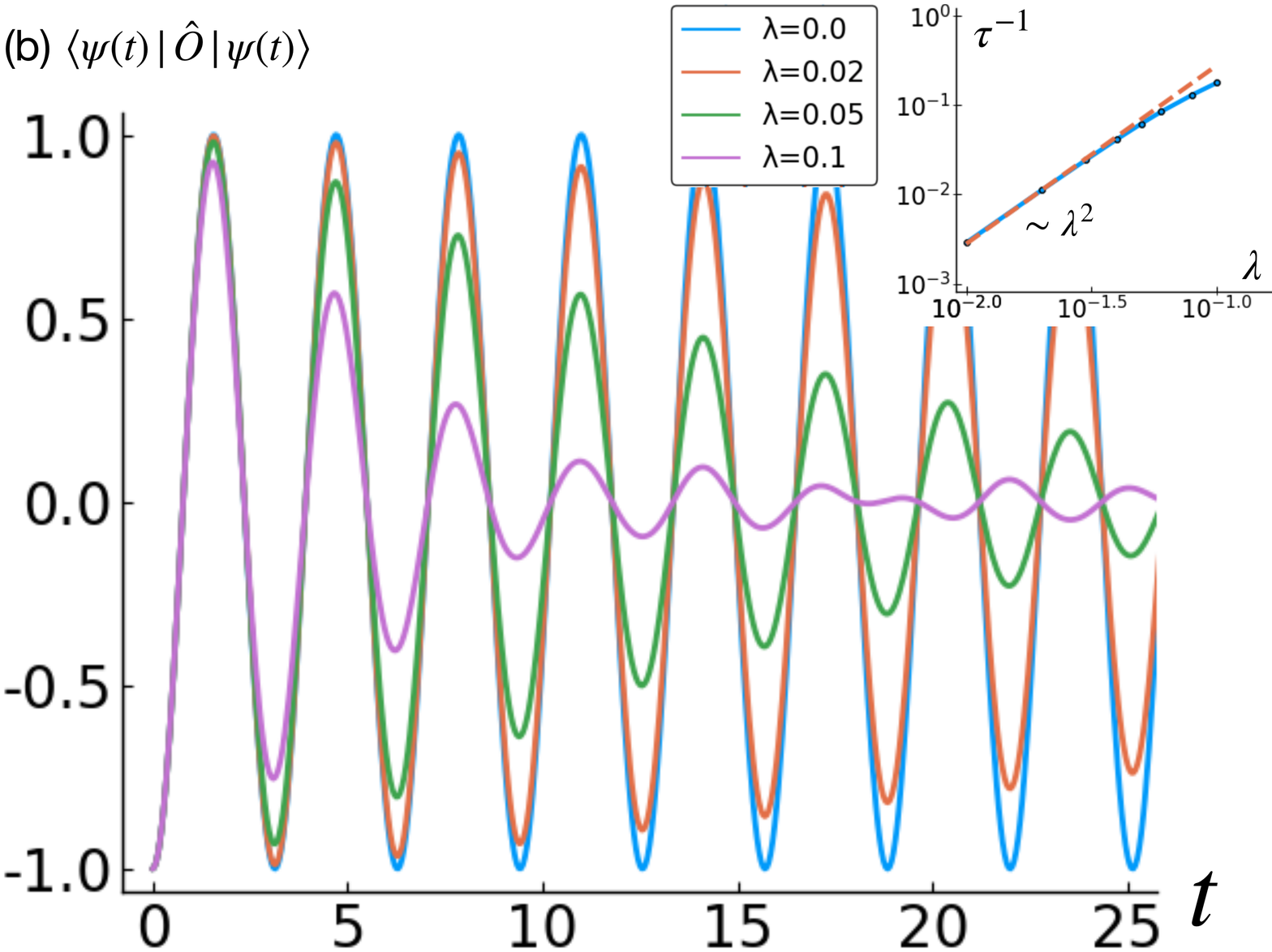}
    \caption{(a) Finite-size scaling of the averaged off-diagonal matrix element $|\la n^{(0)}| \hat{V} |\Gamma \ra|$ within the energy window $|E_{n}^{(0)} - E_{\Gamma}^{(0)}| < 3$, where $|\Gamma \ra$ is chosen as either the scar state $|\mathcal{S}_{m} \rangle$ or a thermal state $|\Gamma_\text{th} \ra$.
    We take $m$ near $L/4$ (more precisely, $m = L/4+1$, $(L-1)/4$, $L/4$, $(L-1)/4$ for $L=12, 14, 16, 18$ respectively, putting the scar state into the $S^z_\text{tot}$ sectors as marked).
    The thermal state is the state with eigenindex equal to the eigenindex of $|\mathcal{S}_{m}\ra$ plus 3.
    (b) The quench dynamics of the state $|\psi_0 \ra = \bigotimes_r \frac{1}{\sqrt{2}}(|+\ra\! +\! (-1)^{r}|-\ra)$ under the perturbed spin-1 XY Hamiltonian $\hat{H} = \hat{H}_0 + \lambda \hat{V}$, with the observable $\hat{O} = \frac{1}{2}[ (S_1^+)^2+(S_1^-)^2]$ in chain of length $L=8$.
    Inset: we extract the decay time by fitting the data to the function $y(t) = A e^{-t/\tau} \cos(\omega t)$, where $A$, $\tau$ and $\omega$ are the fitting parameters.
    } 
    \label{fig:spin1xy}
\end{figure}

In addition to the PXP model, in this appendix we also study effects of a perturbation on the scar states in the spin-1 XY model.
We first briefly review the scar states in this model.
In Ref.~\cite{schecterWeak2019}, Schecter and Iadecola studied the spin-1 XY model: 
\begin{align}
    \hat{H}_0 &= J\sum_{r=1}^L (S^x_r S^x_{r+1} + S^y_r S^y_{r+1}) + h \sum_{r=1}^L S^z_r + D \sum_{r=1}^L (S^z_r)^2 \notag \\
    &+ J_3 \sum_{r=1}^L (S^x_r S^x_{r+3} + S^y_r S^y_{r+3}) ~.
\end{align}
Here we specifically consider a 1D chain with periodic boundary condition.  We denote the local states as $|+\ra$, $|0\ra$, or $|-\ra$, corresponding to the $S^z$ eigenvalue $1$, $0$, or $-1$.
This model has the conserved quantity $S^z_{\text{tot}} = \sum_{r=1}^L S^z_r$ and has the nonthermal eigenstates $|\mathcal{S}_m \ra \equiv \mathcal{N}(m) (J^{+})^m | \Omega \ra$, where $|\Omega \ra = \bigotimes_r |- \ra_r$, the raising operator is $J^+ = \frac{1}{2} \sum_r e^{i \pi r} (S^+_r)^2$, and the normalization factor is $\mathcal{N}(m) = \sqrt{\frac{(L-m)!}{m!L!}}$.
These scar states have energy $E_{|\mathcal{S}_m \ra} = h(2m-L) + DL$ and $S^z_{\text{tot}} = 2m-L$.
We use the same parameters $(J, h, D, J_3) = (1, 1, 0.1, 0.1)$ in $\hat{H}_0$ that were used to demonstrate the above scar states in Ref.~\cite{schecterWeak2019}, see their Fig.2..

We consider the following natural perturbation 
\begin{equation}
    \hat{V} = \sum_{r=1}^L S^z_r S^z_{r+1} ~.
\end{equation}
Therefore, the total Hamiltonian of the perturbed spin-1 XY model is $\hat{H} = \hat{H}_0 + \lambda \hat{V}$. 

First, similar to the analysis in Sec.~\ref{subsec:finitesizescaling}, we examine the averaged off-diagonal matrix element $\la n^{(0)}| \hat{V} |\Gamma \ra$, where $|n^{(0)} \ra$ is an eigenstate of $H_0$ and $|\Gamma \ra$ is a scar state $|\mathcal{S}_m \ra$ or a thermal state $|\Gamma_\text{th} \ra$.
Here we focus on the scar states $|\mathcal{S}_m \ra$ with $m$ near $L/4$ corresponding to the sector $S^z_{\text{tot}}$ near $-L/2$.
The thermal state is chosen as the state with eigenindex equal to the eigenindex of $|\mathcal{S}_m\ra$ plus 3 (we checked that several other choices gave essentially identical results).
In Fig.~\ref{fig:spin1xy}(a), we can see that the averaged off-diagonal matrix element indeed scales as $\sim 1/\sqrt{D}$ for both the scar state and the thermal state, consistent with the finding in the PXP model in Sec.~\ref{subsec:finitesizescaling}.
This also suggest that the scar states $|\mathcal{S}_m \ra$ will eventually thermalize under the perturbation $\hat{V}$ for arbitrary nonzero $\lambda$.

We also examine the effects of the perturbation on the perfect oscillation observed in the XY model for special initial states.
Following Ref.~\cite{schecterWeak2019}, we consider the quench dynamics from the initial state $|\psi_0 \ra = \bigotimes_r \frac{1}{\sqrt{2}}(|+ \ra_r + (-1)^r |-\ra_r)$ and the observable $\hat{O} = \frac{1}{2} [(S_1^{+})^2 + (S_1^{-})^2]$ under the Hamiltonian $\hat{H}$.
At $\lambda = 0$, the dynamics is expected to have perfect oscillation persisting indefinitely.
However, as we increase the perturbation strength $\lambda$, the oscillation starts to dampen, as shown in Fig.~\ref{fig:spin1xy}(b).
The deviation from the perfect oscillation can be rigorously bounded by our theorem Eq.~(\ref{eqn:oscillation_bound}).
We further attempt to extract the thermalization time by fitting the data to $y(t) = A e^{-t/\tau} \cos(\omega t)$, with the fitting parameters $A$, $\tau$, and $\omega$.
In the inset of Fig.~\ref{fig:spin1xy}(b), we show the dependence of $\tau^{-1}$ on $\lambda$.
It appears that at small $\lambda$, we have $\tau^{-1} \sim \lambda^2$ (consistent with the rigorous bound but more reminiscent of the Fermi's golden rule).

\section{A single exact scar state implies nonthermal signatures in some other states}
\label{app:single_scar}
In this appendix, we show that the presence of a single exact scar state actually implies some nonthermalness of some other states nearby in energy.
The argument is inspired by considerations in Sec.~V in Ref.~\cite{moriThermalization2017}.
As an explicit example, we can take the exact scar states in the PXP model $\hH_0$ and consider single-mode approximation (SMA)/multi-mode approximation (MMA) construction of additional approximate scar states in Ref.~\cite{linExact2019}.
However, the argument below is more general and assumes only that correlations of local observables in the scar state are short-ranged.

Let $|\Gamma\rangle$ be an exact scar state of $\hH_0$ with eigenenergy $E_\Gamma$ (the Hamiltonian will remain fixed throughout).
Consider an ``SMA trial state'' of the form
\begin{align}
|\Xi \ra = \sum_j \hat{\xi}_j |\Gamma \ra ~,
\end{align}
where $\hat{\xi}_j$ is a local operator near site $j$.
By subtracting a constant, we can choose $\hat{\xi}_j$ to have zero expectation value in the state $|\Gamma \ra$, which guarantees that $\la \Gamma|\Xi \ra = 0$; we assume this choice throughout.

We now show that the variance of $\hH_0$ in the state $|\Xi \ra$ is finite.
Indeed, we can write
\begin{align}
& (\hH_0 - E_\Gamma) |\Xi \ra = \sum_j [\hH_0, \hat{\xi}_j] |\Gamma \ra ~, \\
& (\hH_0 - E_\Gamma)^2 |\Xi \ra = \sum_j [\hH_0, [\hH_0, \hat{\xi}_j]] |\Gamma \ra ~.
\end{align}
Since $\hH_0$ is a sum of local terms, $\hat{\zeta}_j \equiv [\hH_0, \hat{\xi}_j]$ is a local operator near $j$, and so is $\hat{\eta}_j \equiv [\hH_0, [\hH_0, \hat{\xi}_j]]$.
To calculate the energy variance we need
\begin{align*}
& \la \Xi|\Xi \ra = \sum_{j,j'} \la \Gamma| \hat{\xi}_j^\dagger \hat{\xi}_{j'} |\Gamma \ra = \sum_{j,j'} G_{\xi\xi; \Gamma}(j,j') ~, \\
& \la \Xi| (\hat{H}_0 - E_\Gamma) |\Xi \ra  = \sum_{j,j'} \la \Gamma| \hat{\xi}_j^\dagger \hat{\zeta}_{j'} |\Gamma \ra = \sum_{j,j'} G_{\xi\zeta; \Gamma}(j,j') ~, \\
& \la \Xi| (\hat{H}_0 - E_\Gamma)^2 |\Xi \ra  = \sum_{j,j'} \la \Gamma| \hat{\xi}_j^\dagger \hat{\eta}_{j'} |\Gamma \ra = \sum_{j,j'} G_{\xi\eta; \Gamma}(j,j') ~.
\end{align*}
Remembering our choice $\la \Gamma| \hat{\xi}_j |\Gamma \ra = 0$ and noticing also that $\la \Gamma| \hat{\zeta}_j |\Gamma \ra = \la \Gamma| \hat{\eta}_j |\Gamma \ra = 0$, the above $G_{\xi\xi; \Gamma}(j,j')$, $G_{\xi\zeta; \Gamma}(j,j')$, $G_{\xi\eta; \Gamma}(j,j')$ are connected correlation functions of the corresponding local operators in the state $|\Gamma \ra$.
Using the assumption that $|\Gamma \ra$ has short-ranged correlations, the right-hand-side in each of the above equations is proportional to the system size $L$.
Hence, the variance of $\hH_0$ in the state $|\Xi \ra$ is an $L$-independent number in the limit of large $L$.
As an example, Ref.~\cite{linExact2019} quoted finite variances of the PXP Hamiltonian in the SMA states approximating $E \approx \pm 1.33$ and $\pm 2.66$ scars; these variances were already representative of the thermodynamic limit.

The finite variance of $\hH_0$ in the state $|\Xi \ra$ immediately implies that this state has a non-zero lifetime under the $\hH_0$ dynamics even in the thermodynamic limit.
Note that this is different from a generic trial state whose energy variance in general scales with the system size.
The difference here is that our trial state is actually connected to the exact eigenstate by the action of the sum of local operators.
Since $|\Xi \ra$ is orthogonal to $|\Gamma \ra$, it is a new non-thermalizing state, as we further argue below.

The above prediction of the finite lifetime describes the loss of fidelity upon time-evolving under $\hH_0$ from $|\Xi \ra$ as the initial state.
However, it does not fully capture slow thermalization of observables under such time evolution under the local Hamiltonian.
In fact, we expect the non-thermal properties of the state $|\Xi \ra$ to persist to times that diverge with the system size.

Let us first consider a ``local defect" wavefunction $|x_j \ra \equiv \hat{\xi}_j |\Gamma \ra$.
It is easy to see that the expectation value of a local observable at $j'$ far from $j$ in this state is essentially equal to that in the state $|\Gamma \ra$, where we again use that $|\Gamma \ra$ is short-range-correlated.
Since $|\Gamma \ra$ is a scar state, this initial expectation value is nonthermal.
By the Lieb-Robinson bound, the expectaion value of this local observable in the time-evolved $\exp(-i \hH_0 t) |x_j \ra$ will remain essentially unchanged until time of order $|j - j'|/v_\text{LR}$, where $v_\text{LR}$ is the Lieb-Robinson velocity.
We then conclude that, e.g., until time $L/(4 v_\text{LR})$, half of the sites $j'$ in the system will still have essentially the initial nonthermal value of the local observable.

We can generalize this argument to the state $|\Xi \ra$, which is a coherent superposition of $L$ such local defects $|x_j \ra$ placed at different sites on the lattice.
The time-evolved $|\Xi \ra$ is then a coherent superposition of the time-evolved local defects, and we can apply the above Lieb-Robinson reasoning to each such term.
We then conclude that at a given observation location, e.g., until time $L/(4 v_\text{LR})$, the Lieb-Robinson cone emanating from the defects in half of the terms in the superposition has not reached the observation location. 
In this situation, we expect that the local observable at the observation location is still nonthermal.

We also note that the above reasoning applies also to MMA-type trial states---i.e., multiple applications of the SMA---as long the number of applications is finite.
Together with the possibility of constructing distinct SMA states by using different local defect operators $\hat{\xi}_j$, we thus see that a single scar state indeed implies existence of many additional nonthermalizing low-entanglement trial states nearby in energy.
Note that these trial states have finite energy variance instead of scaling as $O(\sqrt{L})$ but they are not eigenstates.  
It is an interesting open question how the existence of such nonthermalizing trial states is reflected in the actual eigenstates of the Hamiltonian.

Finally, we note that in the context of the PXP model, such SMA/MMA construction~\cite{linExact2019} on top of the analytically known scar states produced competitive approximations to the band of prominent scars found in ED~\cite{turnerWeak2018, turnerQuantum2018}.
One may ask if something like this could happen in other models.
Such SMA/MMA variational states can be constructed in any model with exact scars and should approximate finite-size eigenstates when their energy variance is smaller than the energy level spacing squared.
As the operator used to build ``defects'' atop the exact scar states in this construction can be optimized to minimize the energy variance, it seems likely that at least for small sizes the energy spectrum of a model with exact scar states contains approximate SMA/MMA scar states, even if the model is not proximate to a ``mother model'' where the SMA/MMA states become exact scar states.
For larger sizes, we suspect that the nonthermal aspects of such trial states should still be somehow encoded in the eigenstates over which the trial states can be expanded; however, detecting this nonthermalness in ED could be much more challenging.
It would certainly be useful to test and explore more such ideas in different models with analytic scar states.

%


\end{document}